\documentclass[prc,aps,twocolumn,showpacs,showkeys]{revtex4}
\usepackage{graphicx}
\usepackage{amssymb} 
\begin{document}




\title{Symmetry Energy Effects on the Mixed Hadron-Quark Phase at High
Baryon Density}

\author{M. Di Toro$^{1,2,*}$, B.Liu$^{3,4}$, V.Greco$^{1,2}$, V.Baran$^{5}$
, M.Colonna$^{1}$, S.Plumari$^{1,2}$}

\affiliation{$^{1}$ Laboratori Nazionali del Sud INFN, I-95123 Catania, Italy
\\ 
 $^{2}$ Physics and Astronomy Dept., University of Catania\\
$^{3}$ IHEP, Chinese Academy of Sciences, Beijing, China\\
$^{4}$ Theoretical Physics Center for Scientific Facilities, 
Chinese Academy of Sciences, 100049 Beijing, China\\
$^{5}$ Physics Faculty, Univ. of Bucharest and NIPNE-HH, Romania\\
 $^*$ email: ditoro@lns.infn.it}


\begin{abstract}

The phase transition of hadronic to quark matter at high baryon and 
isospin density is analyzed.
Relativistic mean field models 
are used to describe hadronic matter, and the MIT bag model is adopted
for quark matter. The boundaries of the mixed phase and the related critical
points for symmetric and asymmetric matter are obtained. Due to the different 
symmetry term in the two phases, isospin effects 
appear to be rather significant. 

With increasing isospin asymmetry the binodal 
transition line of the 
($T,\rho_B$) diagram is lowered to a region accessible through heavy ion 
collisions in the energy range of the new planned facilities, e.g. the
$FAIR/NICA$ projects. Some observable 
effects are suggested, in particular an {\it Isospin Distillation} mechanism
with a more isospin asymmetric quark phase, 
to be seen in charged meson yield ratios, and an onset of quark number scaling 
of the meson/baryon elliptic flows
.

The presented isospin effects on the mixed phase appear to be robust
with respect to even large variations of the poorly known symmetry term
at high baryon density in the hadron phase.
The dependence of the results on a suitable treatment of 
isospin contributions in effective QCD Lagrangian approaches,
at the level of explicit isovector parts and/or quark condensates,
is finally discussed.

\end{abstract}

\maketitle

\date{\today}

\begin{figure}
\centering
\includegraphics[scale=0.33]{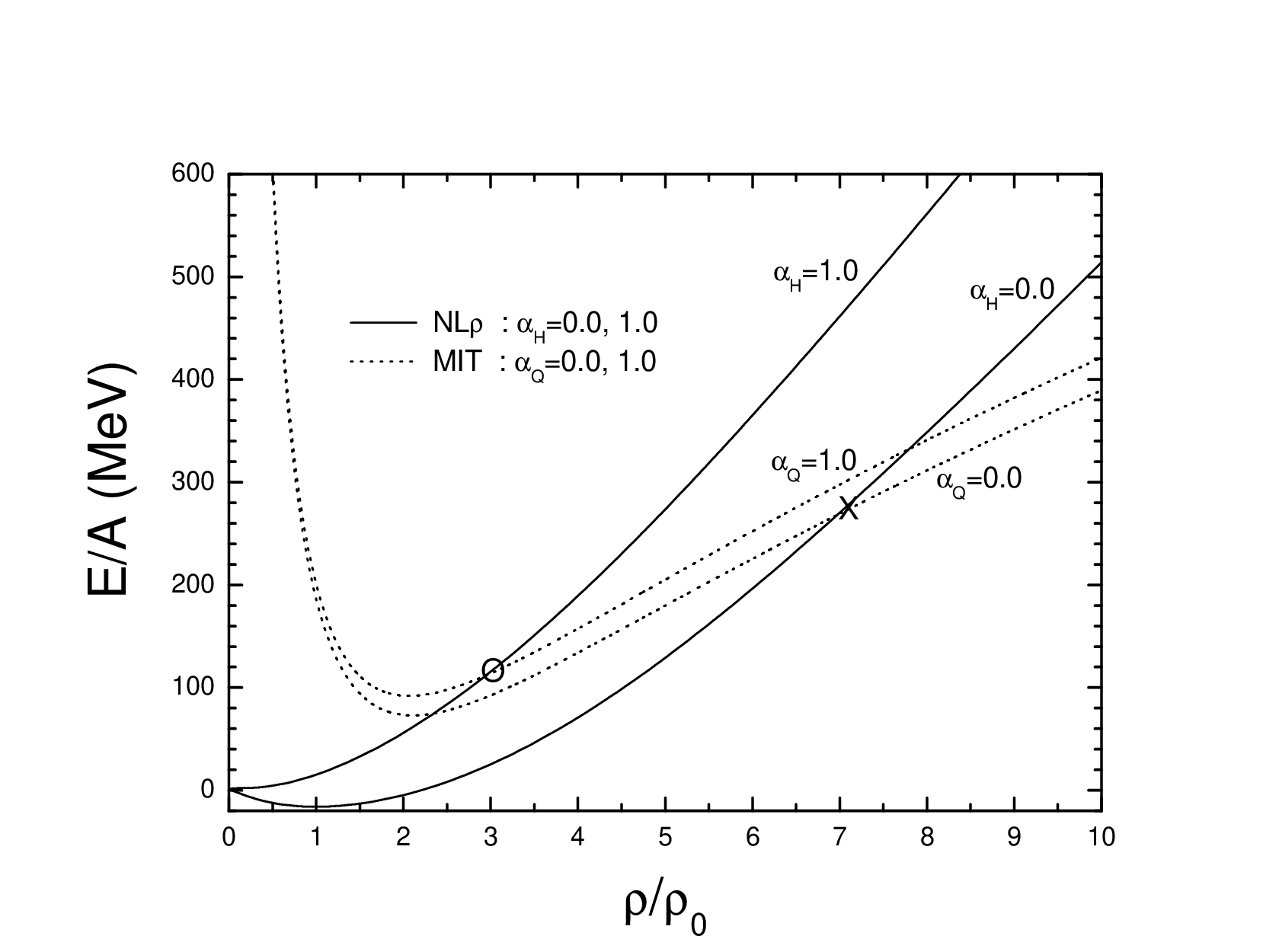} 
\caption{Zero temperature $EoS$ of Symmetric/Neutron Matter: 
Hadron ($NL \rho$), solid lines,
vs. Quark (MIT-Bag), dashed lines. $\alpha_{H,Q}$ represent the isospin 
asymmetry parameters respectively of the hadron,quark matter:
$\alpha_{H,Q}=0$, Symmetric Matter; $\alpha_{H,Q}=1$, Neutron Matter.
}
\label{isoparton} 
\end{figure} 

\section{Introduction}

Several suggestions are already present about the possibility of 
interesting isospin effects on the transition to a mixed hadron-quark phase 
at high baryon density \cite{muller,ditoro,erice08}. This 
seems to be a
very appealing physics program for the new facilities, FAIR at 
GSI-Darmstadt \cite{fair} and NICA at JINR-Dubna \cite{nica}, 
where heavy ion beams (even unstable, with large isospin asymmetry)
will be available with good intensities in the 1-30 AGeV energy region.

The weak point of those predictions is the lack of a reliable Equation of 
State (EoS) that can describe in a consistent way the two phases, 
hadronic and
deconfined, at high baryon density. 

In particular all the Two-EoS models obviously cannot reproduce
continous transitions, like second order phase transitions or cross-overs.
However they can be useful to check if we can have a first-order transition
at lower temperatures. In the latter case, while we cannot localize the 
corresponding Critical End Point, we can study with some confidence the
properties of the mixed phase region if realistic effective interactions 
in the two phases are used. Such discussion will also lead to a
strong  motivation
to work on more refined effective theories for a strong interacting matter.
The aim of our paper is just to show new results, on the dependence on the
EoS choices in the two phases and on possible observables, that would further 
stimulate the search in the field, in theory as well as in experiment.

Isospin effects on the transition are ruled by the symmetry term in the two
phases. For the hadronic side in all the Two-EoS approaches, so far mostly 
applied to develop hybrid models for neutron stars, a rather strong density 
dependence of the symmetry energy has been used
 \cite{muller,ditoro,erice08,burgio02,haensel05,nicotra06,baldo07,burgio08,
providencia09}. This point however is still open mainly due to the present
lack of good data for
isospin effects on Heavy Ion Collisions at intermediate energies, in 
particular on collective flows and particle productions
\cite{baranPR,fuwo06,baoPR,trautmann10,giordano10,ditoro10}. Here we extend
our study also to cases with a much softer hadronic symmetry term in order 
to check the ``robustness'' of the expected isospin effects.

For the quark matter MIT-Bag \cite{MIT}, in refs.
\cite{muller,ditoro,erice08,burgio02,haensel05,nicotra06}, or
Nambu-Jona Lasinio (NJL) \cite{NJL,buballa05}, in refs.
\cite{baldo07,burgio08,providencia09}, models have been adopted, 
always without explicit isospin dependent contributions. Here we also mainly 
used standard MIT-Bag models, but we also discuss the consequence of some 
isospin effects in NJL approaches and possible indirect corrections due to 
the color-pairing residual interaction \cite{pagliara10}.  

We finally like to note that the isospin dependence predictions can be also 
used in the opposite way: if we see such isospin effects on the sensitive 
observables suggested here we can get more confidence on the reliability of 
the used EoS's in the two phases. 

This is the plan of the paper. In Sect.I we present a simple motivation
for a first order transition with isospin effects. In Sect.II the procedure
to construct the ``binodal surface'' from the Gibbs conditions is presented,
 with particular attention to the physical interpretation of the observed
end point. Properties of the mixed phase are evaluated in Sect.III, using
different density dependent symmetry terms for the hadron sector. Sect.IV 
is devoted to the introduction of isospin contributions in the quark 
effective EoS. Sensitive observables in collisions of neutron-rich ions
at intermediate energies are suggested in Sect.V, with relative perspectives.
Finally in Appendix A we present details of the used effective hadron 
interactions and in Appendix B the isospin dependent extension of the NJL model
discussed in the paper.

\section{Why a first-order transition with isospin effects?}

The main qualitative argument in favor of a first order
hadron-quark 
transition at high density and low temperature, with noticeable isospin 
effects,
can be derived from
the Fig.\ref{isoparton}. Here we compare typical Equations of State  for 
Hadron (Nucleon) and Quark 
Matter, at zero temperature, for symmetric 
($\alpha \equiv (\rho_n-\rho_p)/\rho_B \equiv -\rho_3/\rho_B=0.0$) 
and neutron matter 
($\alpha=1.0$), where $\rho_{n,p}$ are the neutron/proton densities and
$\rho_B=\rho_n+\rho_p$ the total baryon density.

In this first simple calculation, a kind of ``homework'', 
for the hadron part we use a Relativistic Mean Field (RMF) EoS
(\cite{SW85,liubo02,baranPR}) 
with non-linear 
terms and an effective $\rho-meson$ coupling for the isovector part, 
largely used 
to study isospin effects in relativistic heavy ion collisions
\cite{erice08,baranPR}. 
However in the paper we will probe several effective hadron interactions
to check the ``robustness'' of the observed symmetry energy effects.
In order to keep a smooth flow of the physics points 
in the discussion, details about the adopted effective nucleon-meson 
Lagrangians
are presented in the Appendix A.

The energy density and the pressure for the quark phase are given by 
the MIT Bag model \cite{MIT} (two-flavor case) and read, respectively:


\begin{eqnarray}\label{edensq}
\epsilon= 3 \times 2 \sum_{q=u,d}\int \frac{{\rm d}^3k}
{(2\pi)^3}\sqrt{k^{2}+m_{q}^{2}}(f_{q}+\bar{f}_{q})
+B~,
\end{eqnarray}

\begin{eqnarray}\label{pressq}
 P =\frac{ 3\times 2}{3}\sum_{q=u,d}\int \frac{{\rm d}^3k}{(2\pi)^3}
\frac{k^2}{\sqrt{k^{2}+m_{q}^{2}}} (f_{q}+\bar{f}_{q})
-B~,
\end{eqnarray}

\noindent
where B denotes the bag constant (the bag pressure), taken as a rather 
standard value from the hadron spectra ($B=85.7~MeV~fm^{-3}$, 
no density dependence), 
$m_{q}$ are the quark masses ($m_u=m_d=5.5~MeV$ choice), 
and $f_q$, $\bar{f}_{q}$ represent
the Fermi distribution functions for quarks and anti-quarks.
The quark number density is given by 

\begin{eqnarray}\label{rhoq}
 \rho_{i}=<q_{i}^{+}q_{i}>={3\times 2} \int \frac{{\rm d}^3k}{(2\pi)^3}
 (f_{i}-\bar{f}_{i})~, ~~~~i=u,d.
\end{eqnarray}

The transition to the more repulsive quark matter will appear around the
crossing points of the two EoS. We see that such crossing for symmetric matter
($\alpha_H=\alpha_Q=0.0$) is located at rather high density, 
$\rho_B \simeq 7\rho_0$, while for pure neutron matter 
($\alpha_H=\alpha_Q=1.0$) it is moving down to about three times
$\rho_0$. Of course the Fig.1 represents just a simple energetic argument to
support the hadron-quark transition to occur at lower baryon densities
for more isospin asymmetric matter. In the rest of the paper we will
rigourously consider the case of a first order phase transition in the Gibbs
frame for a system with two conserved charges (baryon and isospin), in order 
to derive more detailed results. 
Since the first order phase transition presents a jump in the energy, we can 
expect the mixed phase to start at densities even before the crossing points
of the Fig.1. The lower boundary then can be predicted at relatively low 
baryon densities for asymmetric matter, likely
reached in relativistic heavy ion collisions. Moreover this
point is certainly of interest for the structure of the crust and the inner 
core of 
neutron stars, e.g. see refs. 
\cite{burgio02,haensel05,nicotra06,baldo07,burgio08,providencia09} and 
the review 
\cite{page06}. We remark that in ref.\cite{burgio02}  
similar results are obtained
with rather different hadronic approaches, the RMF and the non-relativistic 
Brueckner-Hartree-Fock (BHF) theory. 

We finally note that the above conclusions are rather independent on the 
isoscalar 
part of the used Hadron
EoS at high density, that is chosen to be rather soft in agreement with 
collective flow and
kaon production data \cite{daniel02,fuchs06}.
 
In the used Bag Model no {\it residual} gluon interactions, the 
$\alpha_s$-strong coupling parameter, are 
included. We remark that this in fact would enhance the above effect, 
in the direction of overall lower transition densities, since it
represents an attractive correction for a fixed B-constant, 
see \cite{bmuller95}. A reduction of the Bag-constant with increasing 
baryon density, as suggested by various models, see ref.\cite{burgio02}, 
will also go in the direction of an ``earlier'' (lower density) transition,
as already seen in ref.\cite{ditoro}.
At variance, the presence of
explicit isovector contributions in the quark phase could play an 
important role, 
as shown in the following also for other isospin properties inside the mixed 
phase.

\section{Isospin effects on the Mixed Phase}

We can study in detail the isospin dependence of the transition 
densities
\cite{muller,ditoro,erice08}.
The structure of the mixed phase is obtained by
imposing the Gibbs conditions \cite{Landaustat} for
chemical potentials and pressure and by requiring
the conservation of the total baryon and isospin densities:

\noindent
\begin{eqnarray}\label{gibbs}
&&\mu_{B}^{H}(\rho_B^H,\rho_3^H,T)=\mu_{B}^{Q}(\rho_B^Q,\rho_3^Q,T)~, 
\nonumber \\
&&\mu_{3}^{H}(\rho_B^H,\rho_3^H,T)=\mu_{3}^{Q}(\rho_B^Q,\rho_3^Q,T)~, 
\nonumber \\
&&P^{H}(T)(\rho_B^H,\rho_3^H,T)=P^{Q}(T)(\rho_B^Q,\rho_3^Q,T)~, 
\nonumber \\
&&\rho_{B} = (1-\chi)\rho_B^H + \chi \rho_B^Q ~,
\nonumber \\
&&\rho_{3}= (1-\chi)\rho_3^H + \chi \rho_3^Q ~,
\end{eqnarray}

where $\chi$ is the fraction of quark matter in the mixed phase 
and T is the temperature.

The consistent definitions for the densities and chemical potentials
in the two phases are given by :

\begin{eqnarray}\label{hadronrhomu}
&&\rho_{B}^{H}=\rho_{p}+\rho_{n}, ~~~~\rho_3^H=\rho_p-\rho_n~,
\nonumber \\
&&\mu_B^{H} = \frac{\mu_p + \mu_n}{2}, ~~~~~
\mu_3^{H} = \frac{\mu_p - \mu_n}{2}~,
\end{eqnarray}
for the Hadron Phase and

\begin{eqnarray}\label{quarkrhomu}
&&\rho_B^{Q}=\frac{1}{3}(\rho_{u}+\rho_{d})~,
~~~~~~ \rho_3^{Q}=\rho_{u}-\rho_{d}~,
\nonumber \\
&&\mu_B^{Q}=\frac{3}{2}(\mu_u + \mu_d),~~~~
\mu_3^{Q} =\frac{\mu_u - \mu_d}{2}~,
\end{eqnarray}

for the Quark Phase.

The related asymmetry parameters are:
\begin{equation}
\alpha^{H} \equiv -\frac{\rho_3^{H}}{\rho_B^{H}} =
\frac{\rho_n-\rho_p}{\rho_n+\rho_p}~,~~~~~~
\alpha^{Q} \equiv -\frac{\rho_3^{Q}}{\rho_B^{Q}} =
 3 \frac{\rho_d-\rho_u}{\rho_d+\rho_u}~.
\end{equation}

Nucleon and quark chemical potentials, as well as the pressures in the 
two phases, are directly derived from the respective EoS.

In this way we get the $binodal$ surface which gives the phase coexistence 
region
in the $(T,\rho_B,\rho_3)$ space.
For a fixed value of the
total asymmetry  $\alpha_T=-\rho_3/\rho_B$ 
 we will study the boundaries of the mixed phase
region in the $(T,\rho_B)$ plane. 
Since in general the charge chemical potential is related to the symmetry 
term of the EoS, \cite{baranPR},
$
\mu_3 = 2 E_{sym}(\rho_B) \frac{\rho_3}{\rho_B},
$ 
we expect critical and transition densities
rather sensitive to the isovector channel in the two phases.

In the hadron sector we will use the Non-Linear
Relativistic Mean Field models, \cite{liubo02,baranPR,erice08}, with
different structure of the isovector part, already tested to describe
the isospin dependence of
collective flows and meson production for heavy ion collisions 
at intermediate energies, \cite{greco03,gaitanos04,ferini06}.
We will refer to these different Iso-Lagrangians as: i) $NL$, where no 
isovector
meson is included and the symmetry term is only given by the kinetic 
Fermi contribution, ii) $NL\rho$ when the interaction contribution of
an isovector-vector meson is considered and finally iii) $NL\rho\delta$
where also the contribution of an isovector-scalar meson is accounted for.
 See details in Appendix A1 and refs.\cite{liubo02,baranPR,erice08}.

We will look at the effect on the hadron-quark transition of the different
stiffness of the symmetry term at high baryon densities in the different
parametrizations. As clearly shown in Appendix A1, where a rather transparent
form for the density dependence of the symmetry energy
in RMF approaches is discussed, the potential part of the symmetry term
will be proportional to the baryon density in the $NL\rho$ choice
and even stiffer in the $NL\rho\delta$ case.

We are well aware that there are several uncertainties on the stiffness
of the symmetry energy at high baryon density, mainly due to the lack 
of suitable data, see the reviews \cite{baranPR,baoPR}. Therefore in the
next Section we will show also results with
effective hadron interactions based on RMF models with density
dependent meson-nucleon couplings ($DDRH$ forces, Appendix A2)
that present much softer symmetry terms at high baryon density.
In this way we can directly check the ``stability'' of the observed 
isospin effects on the mixed phase. 

As already mentioned, in the quark phase we use the MIT-Bag Model, where 
the symmetry term is only given by the Fermi contribution. The Bag parameter
B is fixed for each baryon density to a constant, rather standard, value
$B^{1/4}=160MeV$, corresponding to a Bag Pressure of $85.7~MeV~fm^{-3}$.

\begin{figure}
\begin{center}
\includegraphics[scale=0.33]{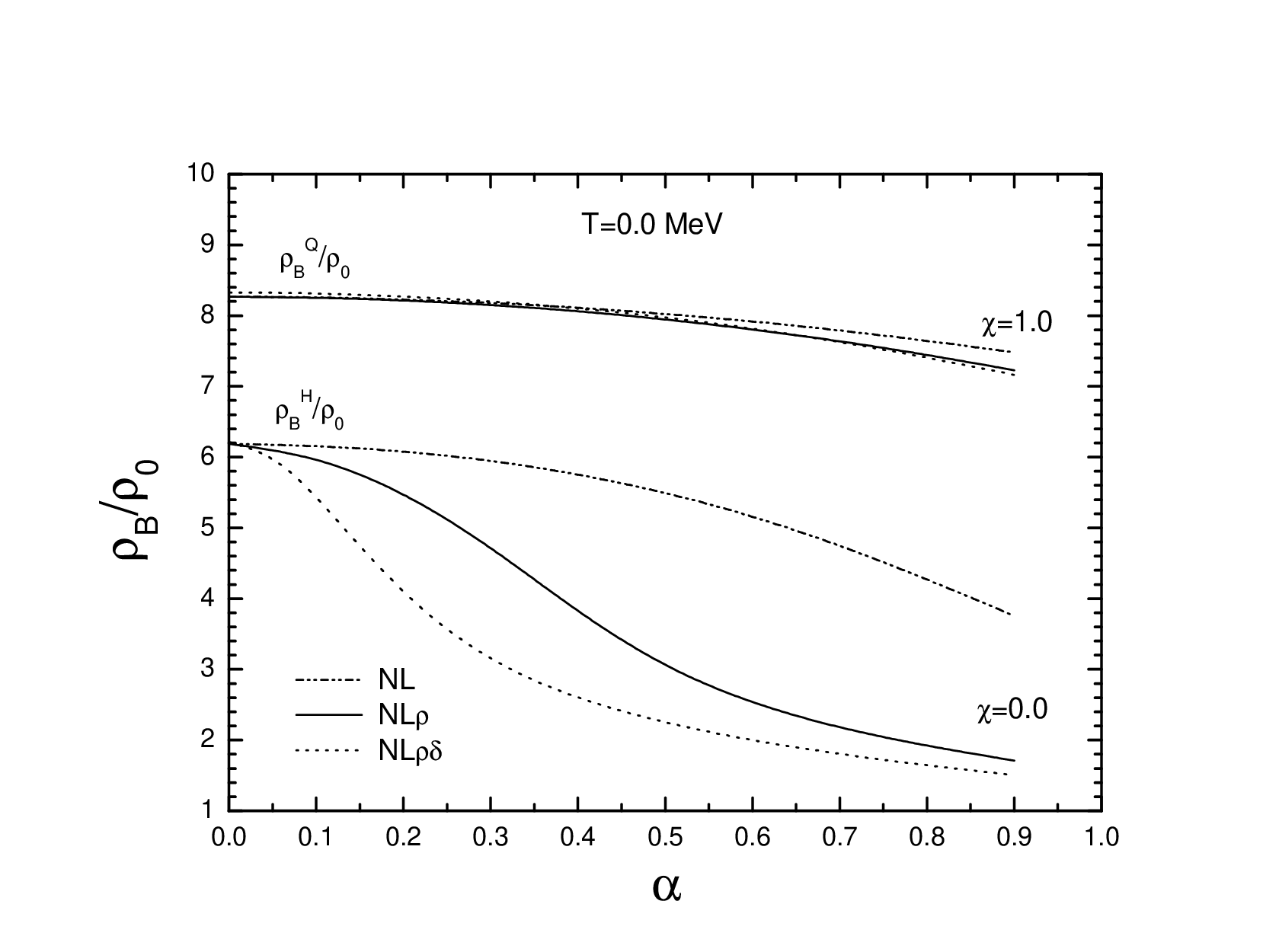}
\caption{
Dependence on the Hadron Symmetry Energy of the Lower ($\chi=0.0$) and
Upper ($\chi=1.0$) Boundaries of the Mixed Phase, at zero temperature,
vs. the asymmetry parameter.
Quark $EoS$: $MIT$ bag model with
$B^{1/4}$=160 $MeV$.
}
\vskip 1.0cm
\label{boundaries}
\end{center}
\end{figure}

In general for each effective interactive Lagrangian we can simulate the 
solution of the highly non-linear system of Eqs.(\ref{gibbs}), via an
iterative minimization procedure, in order to determine the binodal 
boundaries.

A relatively simple calculation can be performed at zero temperature. The 
isospin effect (asymmetry dependence) on the Lower $(\chi=0.0)$ and 
Upper $(\chi=1.0)$ transition densities of the Mixed Phase are shown
in Fig.\ref{boundaries} for various choices of the Hadron EoS. The effect
of a larger repulsion of the symmetry energy in the hadron sector, from
$NL$ to $NL\rho$ and to $NL\rho\delta$,
 is
clearly evident on the lower boundary with a sharp decrease of the transition 
density even at relatively low asymmetries.

\begin{figure}
\begin{center}
\includegraphics[scale=0.33]{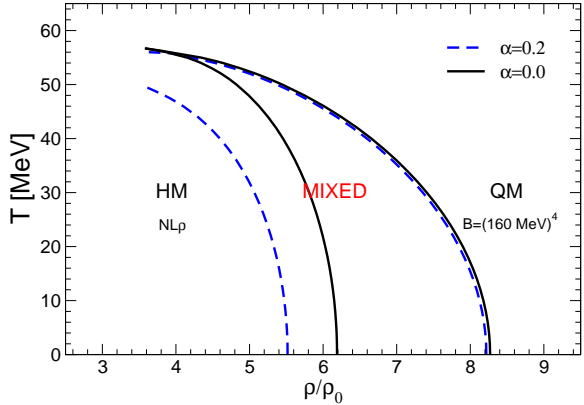}
\caption{
Binodal surface for symmetric ($\alpha=0.0$) and 
asymmetric ($\alpha=0.2$) matter. 
Hadron EoS from
$NL\rho$ interaction. 
Quark $EoS$: $MIT$ bag model with
$B^{1/4}$=160 $MeV$.
}
\label{tcrit}
\end{center}
\end{figure}
 
\begin{figure}
\centering
\includegraphics[scale=0.33,angle=-90]{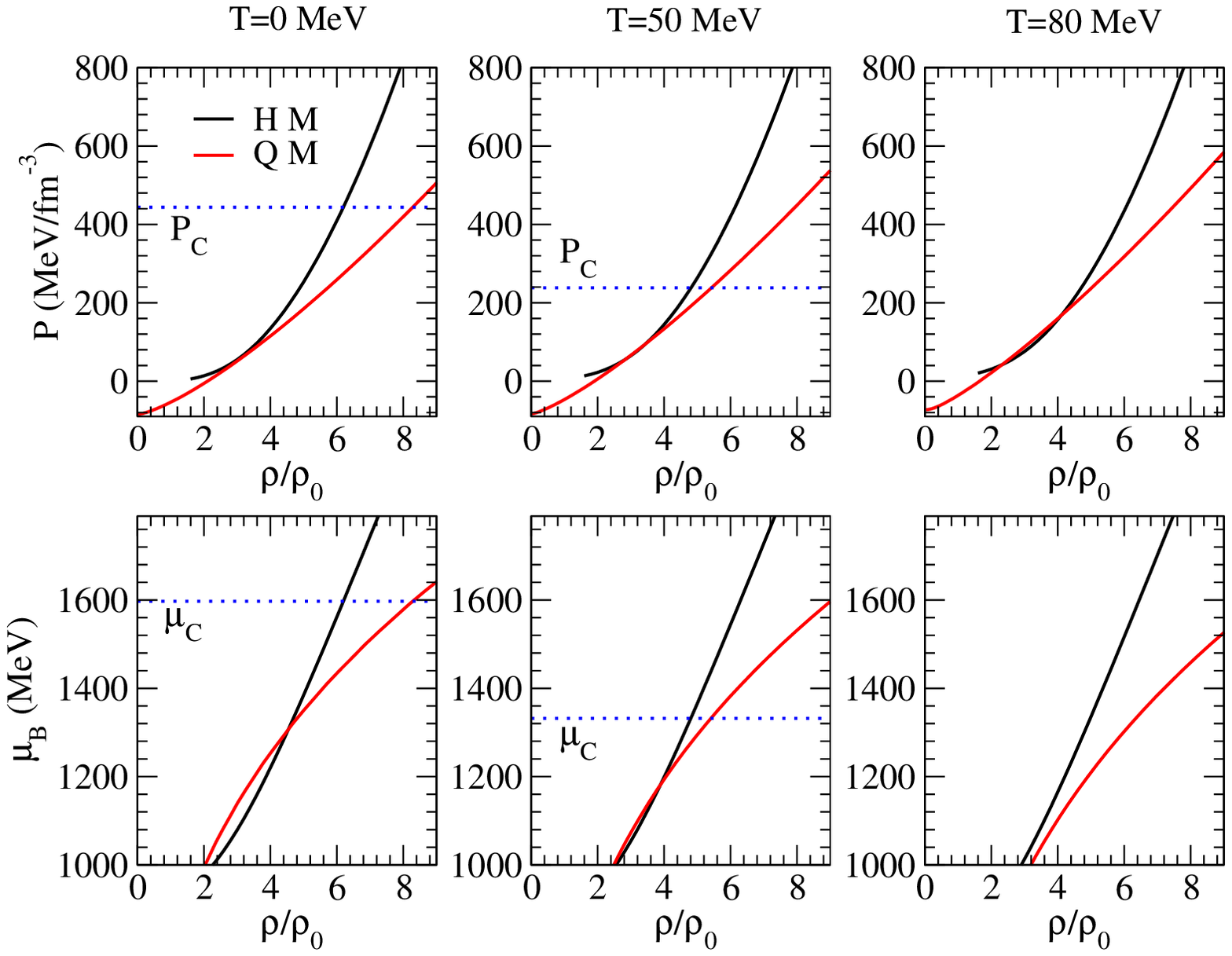}
\caption{
Maxwell construction for symmetric ($\alpha=0.0$) matter at temperatures
T=0, 50 and 80 MeV. The dashed lines correspond to the coexistence values of
pressure (upper panels) and chemical potential (lower panels).
Hadron EoS (black curves) from
$NL\rho$ interaction;  
 Quark $EoS$ (grey curves) from $MIT$ bag model with
$B^{1/4}$=160 $MeV$. 
}
\label{maxwell}
\end{figure}

Typical results for isospin effects on the whole binodal ``surface'' are 
presented 
in Fig.\ref{tcrit} for symmetric and asymmetric matter.
For the hadron part we have started from a $NL\rho$ effective 
Lagrangian
very close to other widely used relativistic effective models, e.g. see
the $GM3$ of ref.\cite{GlendenningPRL18} and the $NL3$ interaction of P.Ring
and collaborators \cite{vretenar03}, which has also given good nuclear 
structure 
results, even for exotic nuclei. 

As expected, the lower boundary
of the mixed phase is mostly affected by isospin effects. In spite of the 
relatively small
total asymmetry, $\alpha=0.2$, we clearly observe in Fig.\ref{tcrit} a shift 
to the left of the
first transition boundary, in particular at low temperature.

In the symmetric matter case the mixed phase is evaluated from the simpler 
Maxwell conditions. The results are shown in Fig.\ref{maxwell} for the 
same hadron and quark $EoS$'s as in Fig.\ref{tcrit} at temperatures T=0, 
50 and 80 MeV. The equal chemical potential densities (intersection of the 
dotted line in the lower panel) must correspond to the equal 
pressure densities of the upper panels.
We nicely see that at T=0 MeV the mixed phase is centered around 
$\rho/\rho_0 \simeq 7.0$ , exactly the $\alpha=0$ crossing point of 
Fig.\ref{isoparton}, confirming our energetic argument about the transition 
location. Precisely the two boundaries are at $\rho_H/\rho_0=6.2$ and
$\rho_Q/\rho_0=8.3$ at a chemical potential $\mu=1597.0 MeV$.

\begin{figure}
\begin{center}
\includegraphics[scale=0.30,angle=-90]{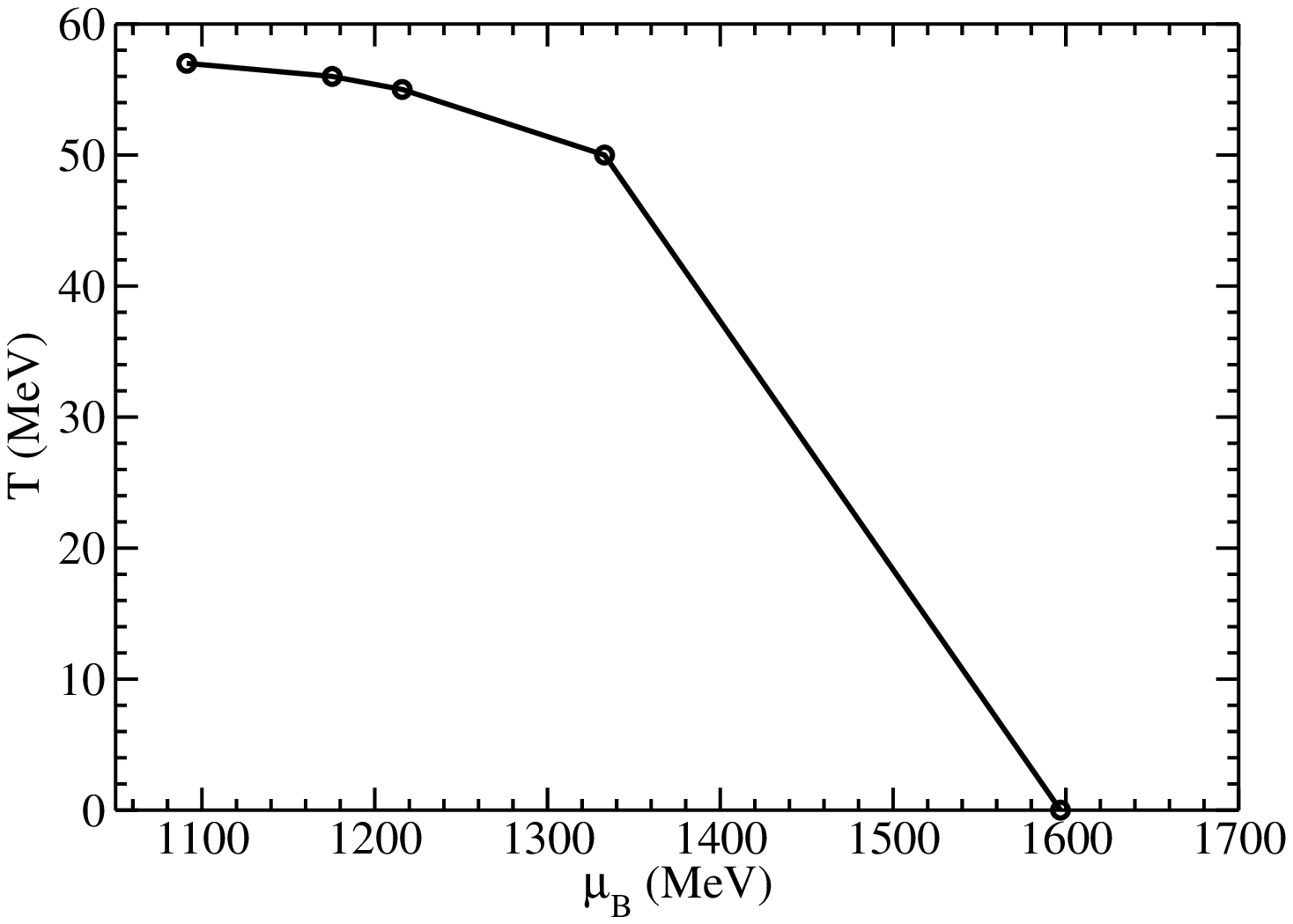}
\caption{
Phase transition line in the $(T,\mu)$ plane for symmetric ($\alpha=0.0$) 
matter. Hadron and Quark EoS like in Fig.\ref{maxwell}. 
}
\label{dtdmu}
\end{center}
\end{figure}
 
\subsection*{About the Critical End Point}

From Fig.\ref{maxwell} we also see that the size of the mixed phase 
is shrinking with temperature, it 
is very narrow at T=50 MeV and finally   
at T=80 MeV 
we cannot have anymore a first order transition. 
In fact a kind of Critical End Point
is appearing at $T_c \simeq 58 MeV$, $\rho_c/\rho_0 \simeq 3.8$, 
$P_c \simeq 120 MeV/fm^3$ and $\mu_c \simeq 1090 MeV$, 
see also Fig.\ref{tcrit}. The result is dependent on the choice of the Bag 
constant, with an increase of the critical temperature with the Bag value due
to the reduction of the pressure in the quark phase, while the chemical 
potentials are not affected. 

However, as already noted in the introduction, within the present Two-EoS 
approach it is
impossible to discuss the nature of the transition around this apparent
Critical End Point. The fact that we reach a point with equal densities in 
the two phases is not implying the onset of a continous transition. 
Indeed from the coexistence conditions of a first order we can have
a point with equal densities but with still a gap in the entropy densities.
Since we can follow the transition in the $(T,\mu)$ plane, such point
will correspond to a zero of the $dT/d\mu$, from the
Clausius-Clapeyron Equation. 

We have checked this possibility for the transition discussed
before, see Fig.\ref{maxwell}, of symmetric matter. 
In Fig.\ref{dtdmu} we present the calculated points of the phase diagram
in the $(T,\mu)$ plane. We see that approaching the
end point of the binodal surface we come very close to the
$dT/d\mu=0$ condition and so we cannot deduce that we have reached a real
Critical End Point of the first order transition.

We note that this result is not implying that the properties of the mixed
phase at lower temperatures discussed within Two-EoS models are
meaningless. We can trust them  if we are using ``realistic'' effective 
interactions in the two 
phases. In fact this is the main point raised here, where the focus is 
on the isospin dependence of the mixed phase
at low temperature, that can be probed in heavy ion collisions at 
intermediate energies.



\begin{figure}
\begin{center}
\includegraphics[scale=0.33]{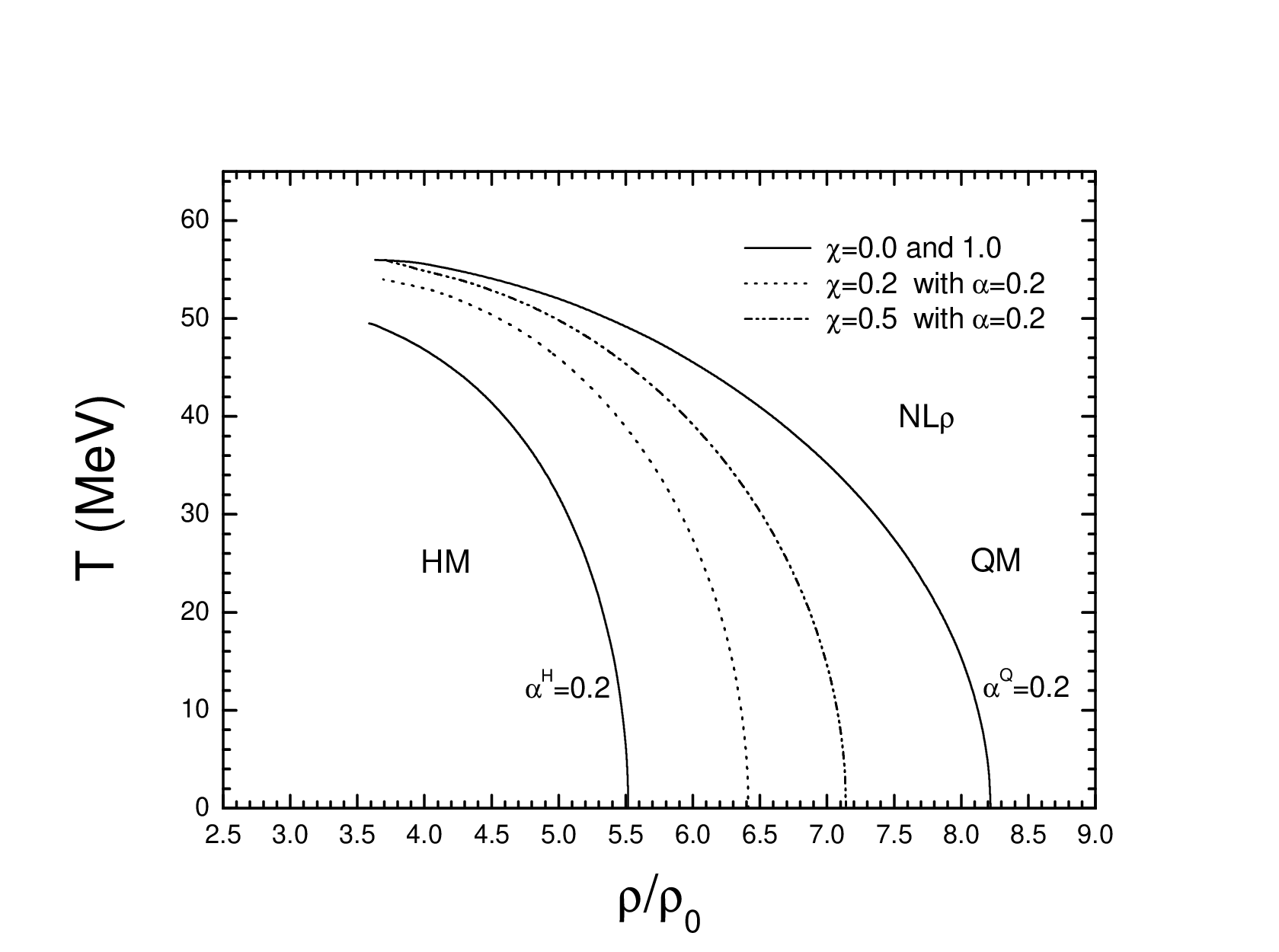}
\caption{Asymmetric $\alpha=0.2$ matter.
Binodal surface and ($T,\rho_B$)
curves for various quark concentrations
($\chi=0.2,0.5$) in the mixed phase.
Quark $EoS$: $MIT$ Bag model with
$B^{1/4}$=160 $MeV$. Hadron $EoS$: $NL\rho$ Effective Interaction.
}
\label{NLrmix}
\end{center}
\end{figure}

\begin{figure}
\begin{center}
\includegraphics[scale=0.33]{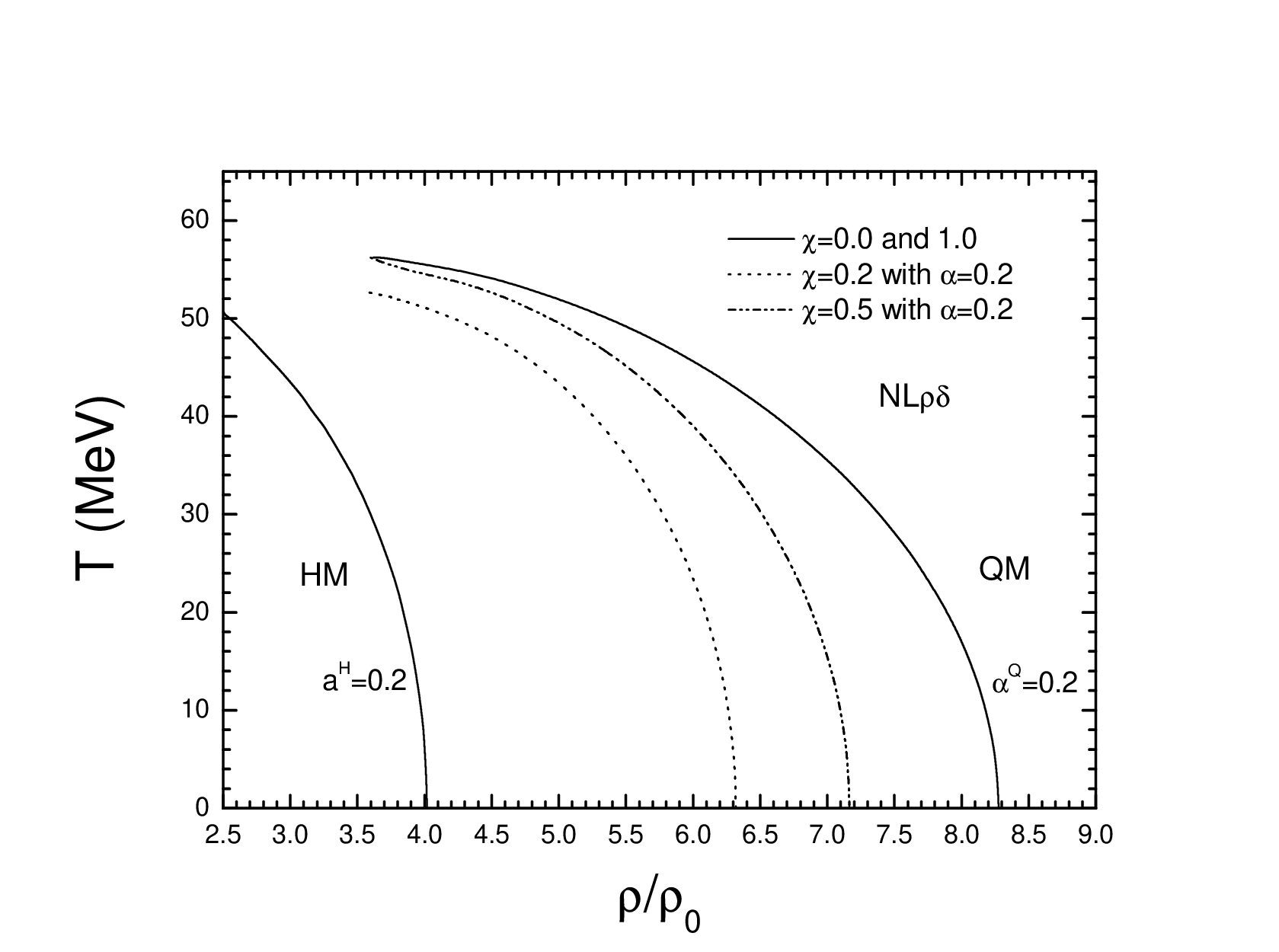}
\caption{As in Fig.\ref{NLrmix}, for the $NL\rho\delta$ Effective Interaction
in the Hadron sector.
}
\label{NLrdmix}
\end{center}
\end{figure}

\section{Inside the Mixed Phase of Asymmetric Matter}

For $\alpha=0.2$ asymmetric matter,
 in the Figs.\ref{NLrmix}, \ref{NLrdmix} we show also the ($T,\rho_B$)
curves inside the Mixed Phase corresponding to a $20\%$ and $50\%$ 
presence of the quark component ($\chi=0.2,0.5$), evaluated respectively
with the
two choices, $NL\rho$ and $NL\rho\delta$, of the symmetry interaction
in the hadron sector. We note, as also expected from Fig.\ref{boundaries},
that in the more repulsive $NL\rho\delta$ case the lower boundary is
much shifted to the left. However this effect is not so evident for the curve
corresponding to a $20\%$ quark concentration, and almost absent for the
$50\%$ case. The conclusion seems to be that for a stiffer symmetry term
in a heavy-ion collision at intermediate energies during the compression stage
we can have more chance
to probe the mixed phase, although in a region with small weight of the
quark component.

\begin{figure}
\begin{center}
\includegraphics[scale=0.33]{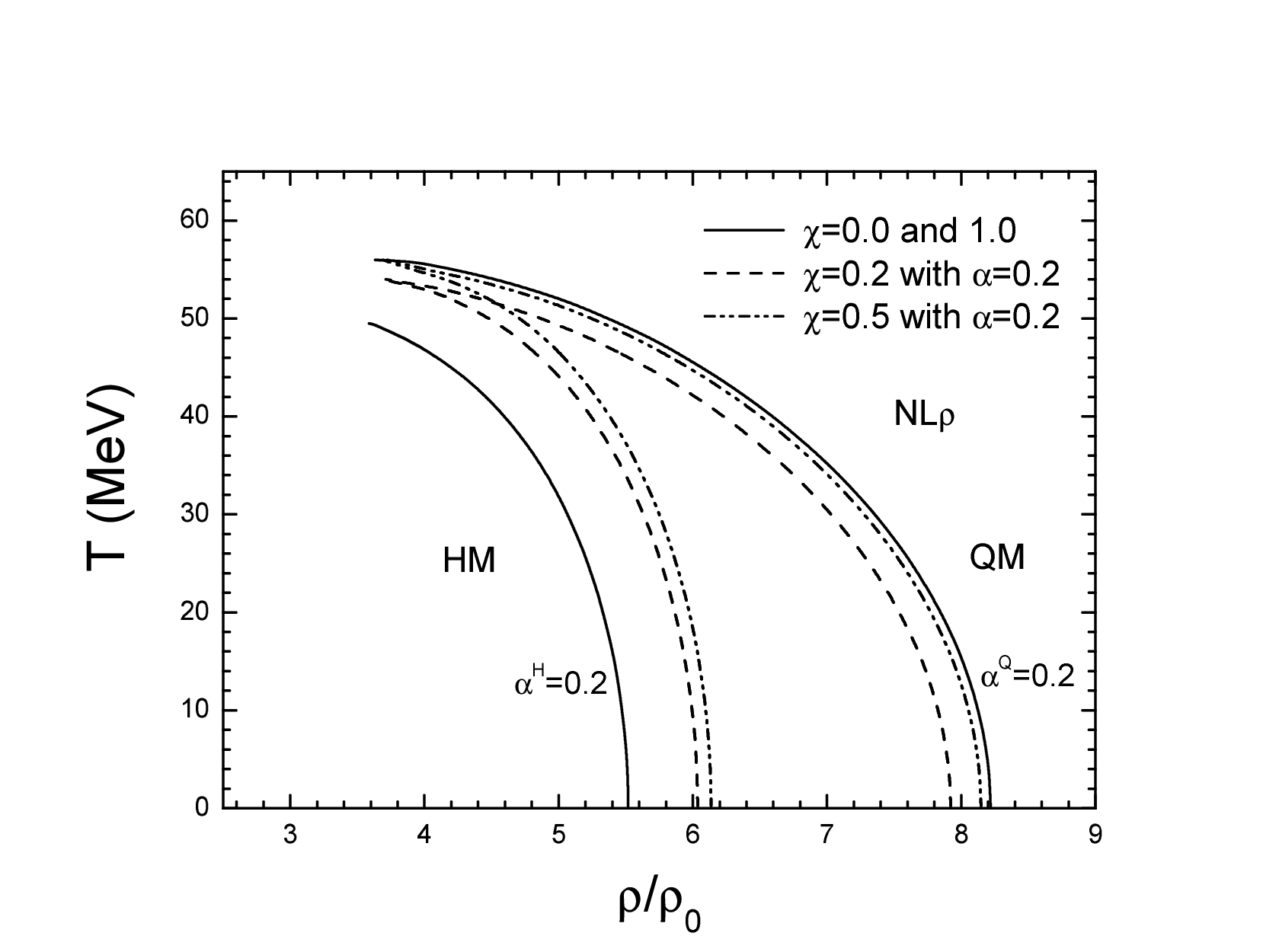}
\caption{Asymmetric $\alpha=0.2$ matter.
Binodal surface and ($T,\rho_B^H,\rho_B^Q$)
curves for various quark concentrations in the mixed phase.
Quark $EoS$: $MIT$ Bag model with
$B^{1/4}$=160 $MeV$. Hadron $EoS$: $NL\rho$ Effective Interaction.
}
\label{NLrmixHQ}
\end{center}
\end{figure}

\begin{figure}
\begin{center}
\includegraphics[scale=0.33]{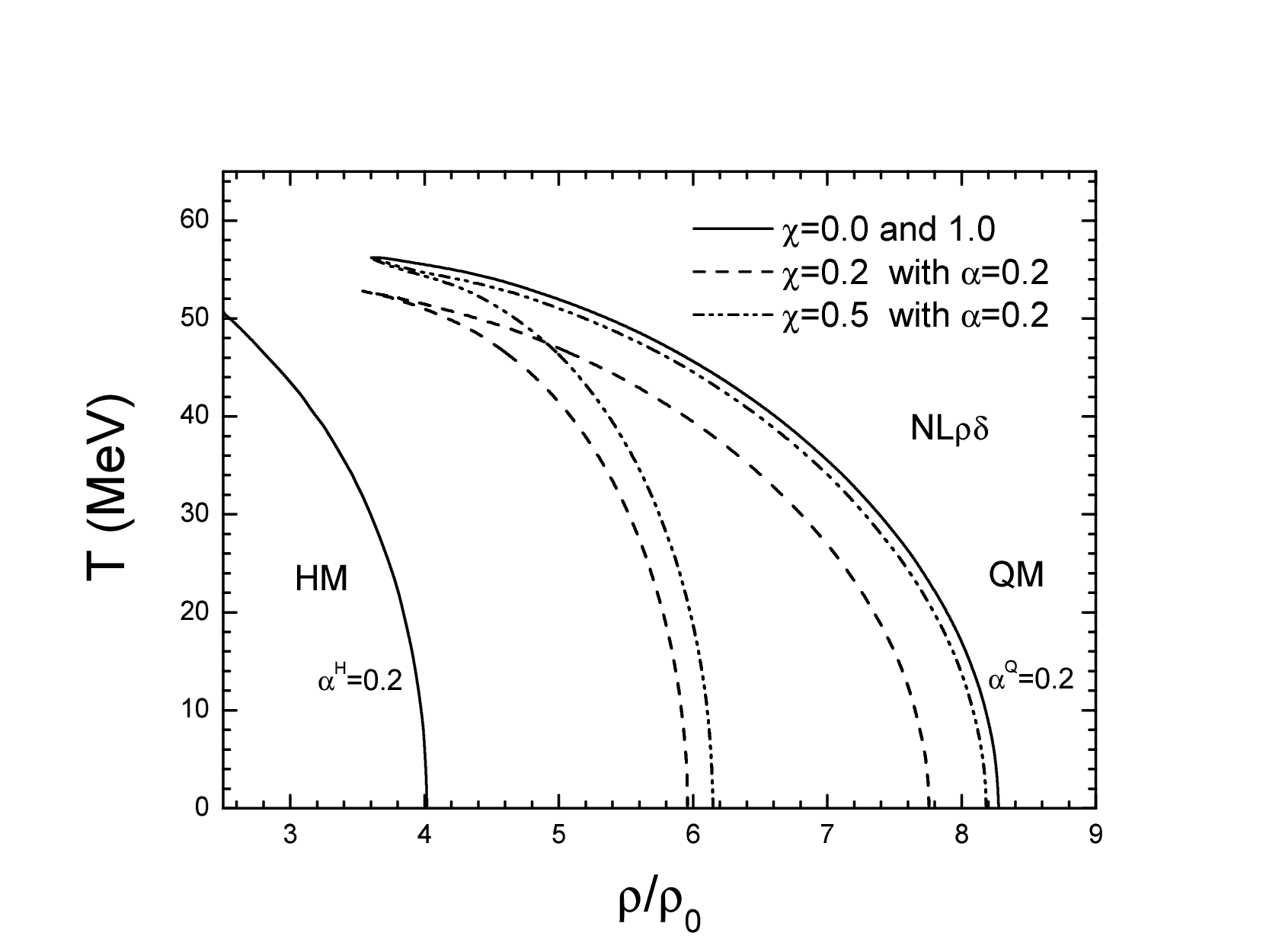}
\caption{As in Fig.\ref{NLrmixHQ}, for the $NL\rho\delta$ Effective 
Interaction in the Hadron sector.
}
\label{NLrdmixHQ}
\end{center}
\end{figure}

In fact from the solution of the system Eq.(\ref{gibbs}) we get 
the baryon densities $\rho_B^H,\rho_B^Q$ in the two phases for any
$\chi$ value. In the Figs. \ref{NLrmixHQ}, \ref{NLrdmixHQ} we present the
results for the same weights $20\%, 50\%$ of the quark phase of the previous 
figures. The quark phase appears always with larger baryon density, even
for the lowest value of the concentration.

Can we expect some signatures related to the subsequent
hadronization in the following expansion?

\begin{figure}
\begin{center}
\includegraphics[scale=0.33]{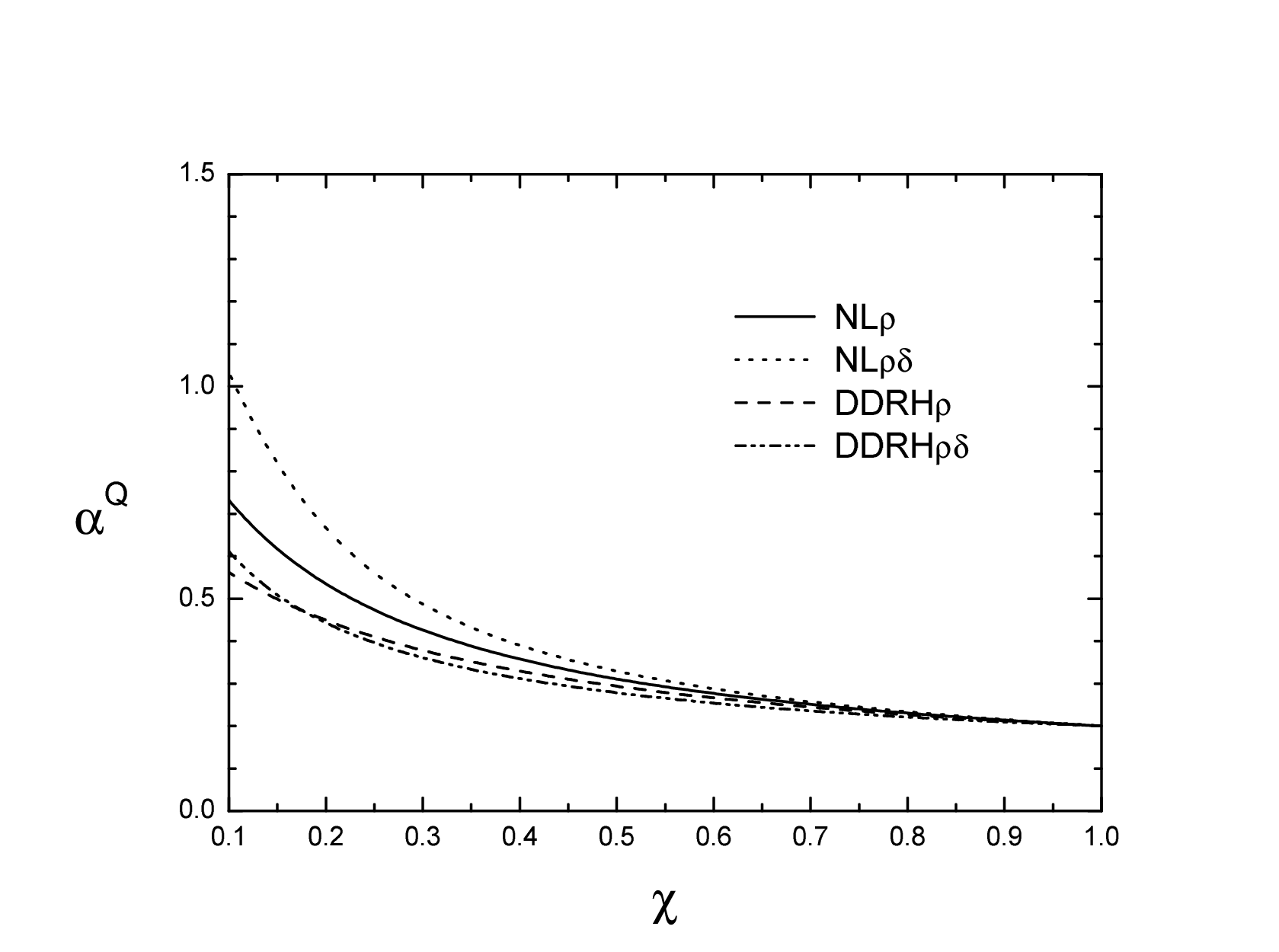}
\caption{
Quark asymmetry in the mixed phase vs. the quark concentration
for asymmetric matter with $T=0$ and $\alpha=0.2$.
$NL\rho$ and $NL\rho\delta$ Effective Hadron Interactions are considered.
The corresponding results with Density Dependent couplings are also shown,
see the Subsection.
Quark $EoS$: $MIT$ bag model with
$B^{1/4}$=160 $MeV$.
}
\label{alphaq}
\end{center}
\end{figure}

An interesting possibility is coming from the study of the asymmetry
$\alpha^Q$ in the quark phase. In fact since the symmetry energy is rather 
different in the two phases we can expect an Isospin Distillation 
(or Fractionation), very similar to the one observed in the Liquid-Gas 
transition in dilute nuclear matter \cite{baran98,chomazPR,baranPR}, this time 
with the larger isospin 
content in the higher density quark phase.

In Fig.\ref{alphaq} we show the asymmetry $\alpha^Q$ in the quark phase
as a function of the quark concentration $\chi$ for the case with global
asymmetry $\alpha=0.2$ (zero temperature). The calculation is performed with 
the two choices of the symmetry term in the hadron sector. We see an impressive
increase of the quark asymmetry when we approach the lower boundary of the
mixed phase, even to values larger than one, likely just for numerical 
accuracy \cite{Qasy}.
Of course the quark asymmetry recovers the global value 0.2 at the upper
boundary $\chi=1$. A simple algebraic calculation allows to evaluate the 
corresponding asymmetries of the hadron phase.
In fact from the charge conservation we have that for any $\chi$-mixture
 the global
asymmetry $\alpha$ is given by:

\begin{equation}\label{alphamix}
\alpha \equiv -\frac{\rho_3}{\rho_B} = 
\frac{(1-\chi)\alpha^H}{(1-\chi)+\chi \frac{\rho_B^Q}{\rho_B^H}} +
\frac{\chi \alpha^Q}{(1-\chi) \frac{\rho_B^H}{\rho_B^Q}+ \chi}
\end{equation}

For any $\chi$, from the calculated $\alpha^Q$ of Fig.\ref{alphaq}
and the $\rho_B^H,\rho_B^Q$ of Figs.\ref{NLrmixHQ}, \ref{NLrdmixHQ},
we can get the correspondent asymmetry of the hadron phase $\alpha^H$.
 For a $20\%$ quark concentration
we have an $\alpha^Q/\alpha^H$ ratio around 5 for $NL\rho$ and around 20
for $NL\rho\delta$, more repulsive in the isovector channel. 
It is also interesting to compare the isospin content $N/Z$ of the high
density region expected from transport simulations without the Hadron-Quark
transition and the effective $N/Z$ of the quark phase in a 
$20\%$ concentration.
In the case of $Au+Au$ (initial $N/Z=1.5$) central collisions at $1AGeV$
in pure hadronic simulations we get in the high density phase a reduced
$N/Z \sim 1.2-1.25$ (respectively with $NL\rho\delta-NL\rho$ interactions)
due to the fast neutron emission \cite{gaitanos04,ferini06}. The corresponding
isospin content of the quark phases is much larger, $N/Z=3.0$ for $NL\rho$ 
and $N/Z=5.7$ for $NL\rho\delta$. This is the $neutron~trapping$ effect
discussed in the Section 5.
We would expect
a signal of such large asymmetries, coupled to a larger baryon density in 
the quark phase, in the subsequent 
hadronization.

We finally remark that at higher temperature and smaller baryon chemical 
potential (ultrarelativistic collisions) the isospin effects discussed here 
are going to vanish \cite{sissakian08}, even if other physics can enter 
the game and charge asymmetry effects are predicted also at $\mu_B=0$ and
$T \simeq 170~MeV$ \cite{kogut04,toublan05}.

\subsection*{Results with a softer symmetry term at high baryon densities}

In order to account for the the present uncertainties on our knowledge 
of the symmetry term of the hadron EoS at high baryon density, see also the 
recent \cite{giordano10}, we have performed a new calculation using
a RMF hadron interaction which gives a much softer behavior of the 
symmetry energy at high densities. In this way we can check the ``robustness''
of the expected isospin effects on the mixed phase discussed before.
We use a Density Dependent Relativistic Hadron (DDRH) field approach,
where an explicit density dependence of the meson-nucleon couplings
is introduced \cite{fuchs95,typel99,avancini04}, see details in Appendix A2.
As clearly shown in Fig.\ref{esymDD}, 
 the main difference with respect to the previously presented results is 
that the symmetry energy is now less repulsive at high density.
This is due to the fact that, 
following some indications from Dirac-Brueckner calculations
\cite{hofmann01,goegelein08}, the isovector-meson couplings 
in the $DDRH\rho$ and $DDRH\rho\delta$ cases show an
increase for the attractive $\delta$-field and a decrease for the
repulsive $\rho$-field, see Fig.\ref{couplings} in Appendix A2.

\begin{figure}
\begin{center}
\includegraphics[scale=0.33]{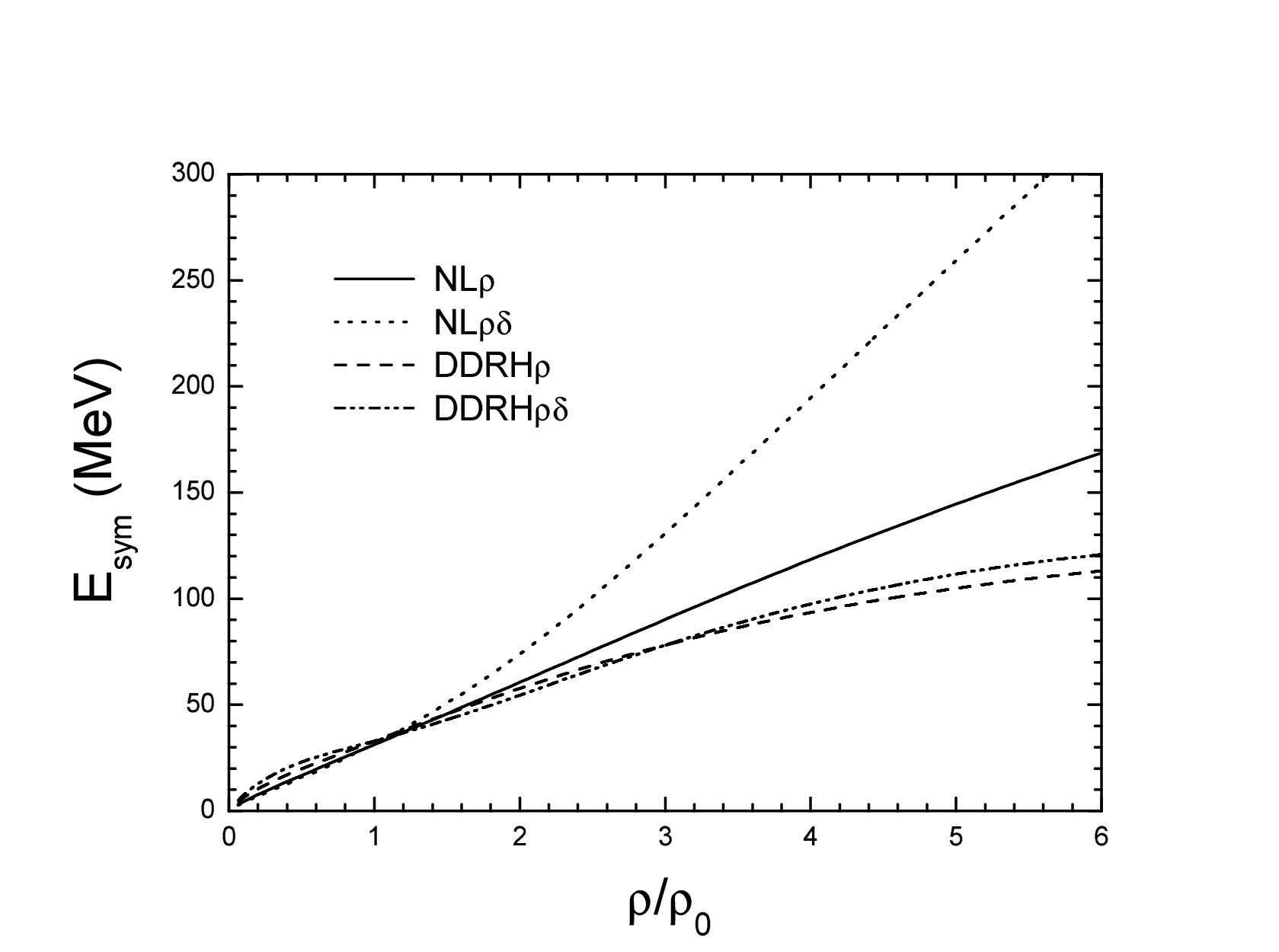}
\caption{
Density dependence of the symmetry energy for the used RMF hadron
effective models.
$NL\rho$ and $NL\rho\delta$ represent the Non-Linear Effective Hadron 
Interactions with constant couplings.
The corresponding results with Density Dependent couplings are also shown.
}
\label{esymDD}
\end{center}
\end{figure}

Moreover interesting rearrangement terms are
now present in the pressure and in the baryon chemical potentials, 
proportional to the density slopes of the 
couplings (see Appendix A2) and so particularly important at high densities,
 as also shown in neutron star applications \cite{liubo07}. 
 
\begin{figure}
\centering
\includegraphics[scale=0.33]{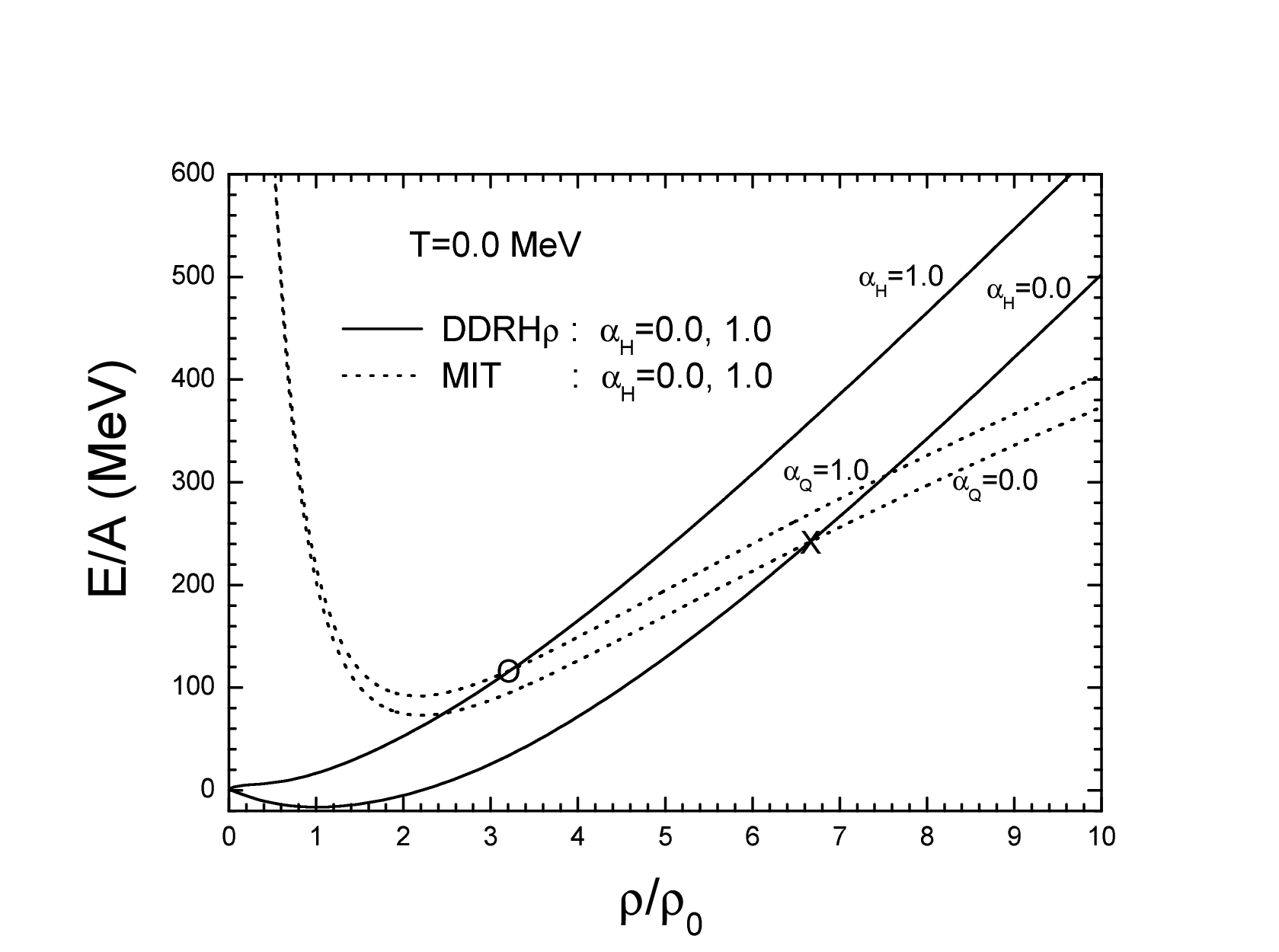} 
\caption{Zero temperature $EoS$ of Symmetric/Neutron Matter: 
Hadron ($DDRH\rho$), solid lines,
vs. Quark (MIT-Bag), dashed lines. $\alpha_{H,Q}$ represent the isospin 
asymmetry parameters respectively of the hadron,quark matter:
$\alpha_{H,Q}=0$, Symmetric Matter; $\alpha_{H,Q}=1$, Neutron Matter.
}
\label{isopartonDD} 
\end{figure}

\begin{figure}
\begin{center}
\includegraphics[scale=0.33]{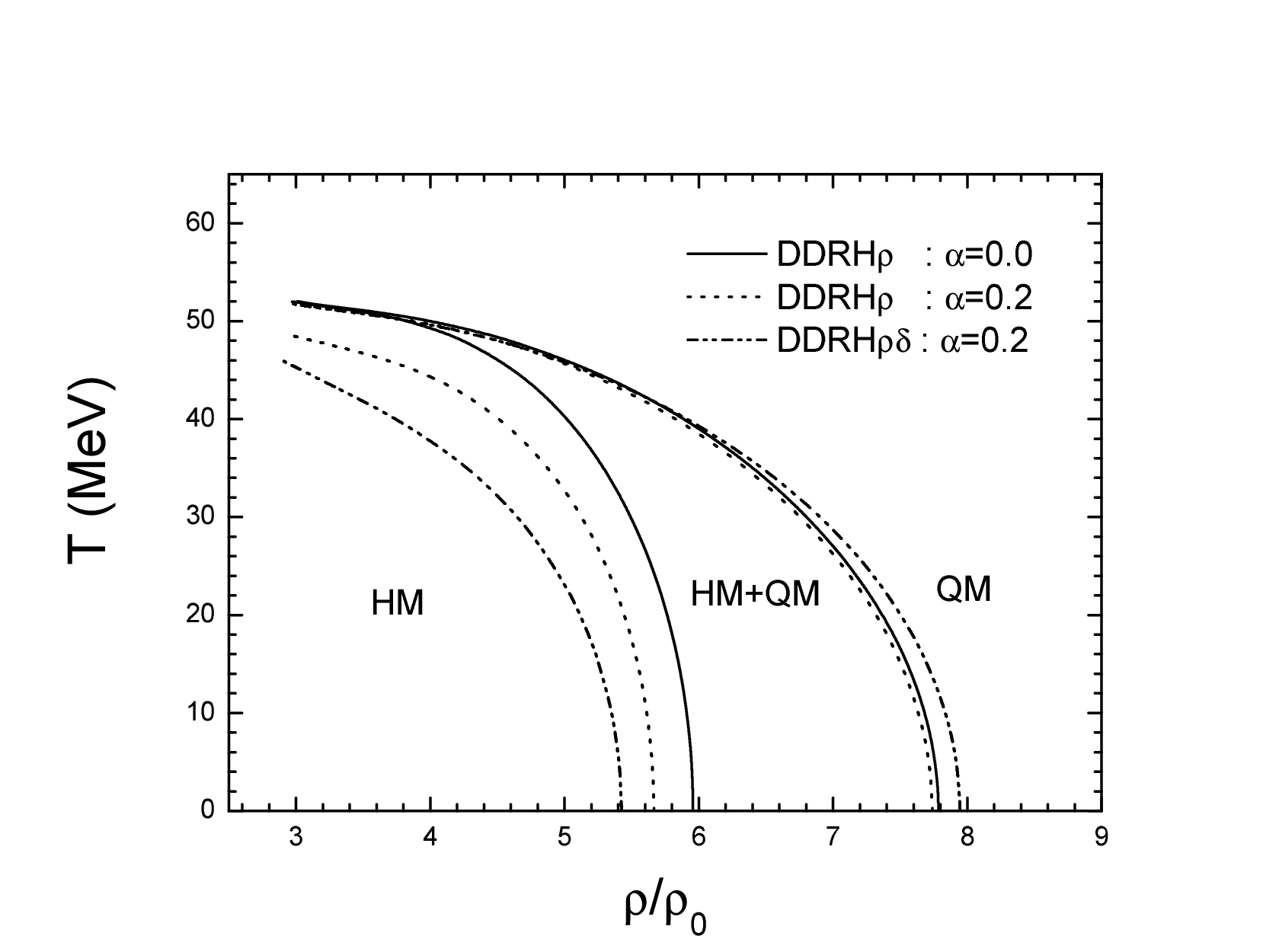}
\caption{
Binodal surface for symmetric ($\alpha=0.0$) and 
asymmetric ($\alpha=0.2$) matter. 
Hadron EoS from
$DDRH$ interactions. 
Quark $EoS$: $MIT$ bag model with
$B^{1/4}$=160 $MeV$. 
}
\label{tcritDD}
\end{center}
\end{figure}

We present first some results on the shift to lower densities of the onset 
of the mixed phase with increasing isospin asymmetry.
In Fig.\ref{isopartonDD} we have the result with $DDRH\rho$ 
supporting the crossing 
argument of the Fig.\ref{isoparton}. Fig.\ref{tcritDD} shows in more detail
the shift to the left of the lower boundaries. The curves should be compared 
to the corresponding lines of the $NL$-constant coupling model:
$DDRH$ to the solid lines of Fig.\ref{tcrit} ($NL$, $\alpha=0.0$),
 $DDRH\rho$ to the solid lines of Fig.\ref{NLrmix} ($NL\rho$, $\alpha=0.2$)
and finally $DDRH\rho\delta$ to the solid lines of Fig.\ref{NLrdmix}
 ($NL\rho\delta$, $\alpha=0.2$). 
We see that the isospin effects of the
hadron-quark transition are still present, although with some reduction.

Finally the new $Isospin~Distillation$ effects are shown as the $DDRH$
results added in Fig.\ref{alphaq}, about the isospin asymmetry in the 
quark phase for different quark concentrations. We note that for $20\%-30\%$
quark components we still see a noticeable increase of the isospin asymmetry.

We can conclude that the revealed isospin asymmetry effects on the
hadron-quark mixed phase at high baryon density appear to be rather 
``robust''
with respect to relatively large variations of the stiffness of 
the symmetry term.

\section{Isospin in Effective Quark Models}

All the above results will be also sensitive to the explicit
inclusion of isovector interactions in effective 
non-perturbative QCD models at high baryon chemical potentials.
Unfortunately few attempts have been worked out for two main reasons:
i) the difficulties of lattice-QCD calculations at high baryon densities;
ii) the main interest on the QGP phase transition at high temperature and
small baryon chemical potentials, as probed in the expanding fireball
of ultrarelativistic heavy ion collisions.
A first approach can be supplied by a two-flavor Nambu-Jona Lasinio ($NJL$) 
model \cite{NJL}, which in fact describes the chiral restoration but not
the deconfinement dynamics. 
The isospin asymmetry can be included in a flavor-mixing picture 
\cite{frank03,buballa05}, corresponding to different couplings to
the (u,d) quark-antiquark condensates.
As a consequence we can have now a 
dependence of the constituent mass of a given flavor to both 
quark condensates.
We devote the Appendix B to a detailed study of this isospin effects in the
$NJL$ chiral dynamics. 

Due to the scalar nature of the interacting part of the
corresponding Lagrangians only the quark effective mass dynamics will 
be affected.
In the ``realistic'' small mixing 
case, see also \cite{frank03,shao06}, we get a definite $M_u^*>M_d^*$ 
splitting at high baryon density (before the chiral restoration). 

All that can indicate a more fundamental confirmation of the $m^*_p>m^*_n$ 
splitting in the hadron phase,
as suggested by the effective $QHD$ model with the isovector scalar $\delta$ 
coupling, see \cite{liubo02,baranPR}.
However such isospin mixing effect results in a very small variation
of the symmetry energy in the quark phase, related only to the Fermi
kinetic contribution. 
Moreover we remind that confinement 
is still 
missing in this $NJL$ mean field approach. In any case there are extensive
suggestions about a favored chiral symmetry restoration in systems with
large neutron excess \cite{kaiser09}.

More generally starting from the QCD Lagrangian one can arrive to an 
effective color current-current interaction where an expansion in various 
components can provide isovector contributions, \cite{weise07}.

In this respect we remark another interesting ``indirect'' isospin effect,
i.e. not directly coming from isovector terms in the effective Lagrangian,
but
related to the presence of quark condensates due to the attractive gluon
interaction. We note that
just a color-pairing mechanism in the two-flavor
system (the $2SC$ phase \cite{alford08}) would imply a stiffer symmetry energy
in the quark EoS since we have a larger attraction when the densities of up 
and down quarks are equal. A first study of the high density hadron-quark 
transition including such gluon correlation in the Bag model has been 
presented very recently in \cite{pagliara10}. 
Now the symmetry energy difference between hadron and quark phases is
partially reduced, at least at low temperatures, and consequently also
the isospin effects discussed in detail in this work will be weaker,
although still present. An interesting point is that in any case the
quark phase is more bound due to the attractive gluon contribution.
Hence the transition to 
the mixed phase will still appear at relatively low baryon densities,
 now for an ``isoscalar'' mechanism,  
within the reach of ``low energy''
heavy ion collisions, i.e. in the range of few AGeV. 
As an intuitive picture we can refer again to the Fig.1. Essentially
the difference between the $\alpha_Q=0.0$ and $\alpha_Q=1$ curves
is increasing but meanwhile both are decreasing.

With increasing temperature the color pairing effect will be in general 
reduced, as confirmed in \cite{huang03} in an extended $NJL$ calculation,
and so isospin effects, as discussed before, will be more relevant. All that
 is naturally related to the used value of 
the superconducting gap,
opening new stimulating perspectives.
In this sense new experiments on mixed phase properties observed with
isospin asymmetric heavy ion collisions, as suggested in the final section,
will be extremely important.

\section{Perspectives and suggested observables}

Based on the qualitative argument of the Introduction and on more detailed
calculations in a first order phase transition scheme, we have predicted
rather ``robust'' isospin effects on the hadron-quark transition at
high baryon densities, not depending on details of the EoS
parametrizations in the hadron and quark phases.

Our results  seem to indicate a specific region where the onset of the
mixed phase should be mainly located: $2<\rho_B/\rho_0<4$, $T \leq 50-60~MeV$,
 for realistic asymmetries $\alpha \sim 0.2-0.3$.
A key question is if such a region of the phase space can be explored by means
of Heavy-Ion-Collisions. In refs.\cite{ditoro,erice08} it is shown that even
collisions of stable nuclei at intermediate energies ($E/A \sim 1-2~GeV$)
make available the pertinent ($T,\rho_B,\alpha$) region where the
phase transition is expected to occur.


In this respect we can refer to the
reaction $^{238}U+^{238}U$ (average isospin asymmetry $\alpha=0.22$) at
$1~AGeV$ that has been investigated in ref.\cite{ditoro}, using
a consistent Relativistic Mean Field approach with the same interactions,
  for a semicentral 
impact parameter $b=7~fm$, chosen  
just to increase
the neutron excess in the interacting region. 
The evolution of momentum distribution
and baryon/isospin densities in a space cell located in the c.m. of the 
system has been also studied.
After about $10~fm/c$ a local
equilibration is achieved still in the compressed phase, before 
the fast expansion.  We have a unique Fermi distribution and
from a simple fit the ``local'' temperature can be evaluated. 
A rather exotic nuclear matter is formed in a transient
time of the order of $10-20~fm/c$, with baryon density around $3-4\rho_0$,
temperature $50-60~MeV$, and isospin asymmetry
between $0.2$ and $0.3$, likely inside the estimated mixed 
phase region.

Of course a relatively higher beam energy will allow
to enter more deeply into the mixed phase. Such energies
will however be available in the next future. In particular
we notice that
 high intensity $^{238}U$ beams in this energy range
would be delivered in the first stage of the FAIR facility
\cite{fair,horst} and also at JINR-Dubna in the Nuclotron
first step of the NICA project \cite{sissakian08}.

Which are the observable effects to look at if we enter and/or cross the 
mixed phase?

As already stressed, a first expectation will be the Isospin Distillation 
effect, a kind of {\it neutron trapping} in the quark phase, supported by 
statistical fluctuations   \cite{ditoro} as well as by a 
symmetry energy difference in the
two phases, as  discussed in Section III.
In fact while in the pure hadron matter (neutron-rich) at high density 
we have a large neutron
potential repulsion (in both $NL\rho$, $NL\rho\delta$ as well as
in the corresponding $DDRH$ cases), in the
quark phase the $d$-quarks see a smaller symmetry repulsion 
essentially only
due to the kinetic contribution 
from the Fermi gas.
As a consequence while in a
pure hadronic phase neutrons are quickly emitted or ``transformed'' in
protons by inelastic collisions \cite{ferini06}, when the mixed phase
starts forming, neutrons are kept in the interacting system,
in the quark phase, where they can even thermalize, up to the
subsequent hadronization in the expansion stage \cite{erice08}.
Observables related to such neutron ``trapping'' could be
\begin{itemize}
\item{(i). An
inversion in the trend of emission of fast neutron rich clusters
with increasing beam energy, 
to be seen in the $n/p,~^3H/^3He..$ ratios at high kinetic energies;}
\item{(ii). 
An enhancement of the production of isospin-rich nucleon resonances and 
subsequent decays, that can be evaluated via equilibrium statistical
approaches \cite{ferini05};}
\item{(iii). Related to the previous point,
 an increase of $\pi^-/\pi^+$, $K^0/K^+$ yield ratios 
for mesons coming from high density regions, to be selected via
large transverse
momenta, corresponding to a large radial flow.}
\end{itemize} 
 
 If such kinetic selection of particles from the mixed phase can 
really be successful also other potential signatures would become available.
One is related to the general softening of the matter, due to the contribution
of more degrees of freedom, that should show up in the damping of collective 
flows \cite{csernai99}.

The azimuthal distributions (elliptic flows) will be
particularly affected since particles mostly retain their high transverse
momenta escaping along directions orthogonal to the reaction plane without
suffering much rescattering processes. 
Thus a further signature could be the observation, for the selected particles, 
of 
the onset of a quark-number scaling of the elliptic flow: a property of 
hadronization
by quark coalescence that has been predicted and observed at RHIC energies,
i.e. for the transition at $\mu_B=0$ \cite{fries08}.

We  note that all the above results, on the Binodal Boundaries of the
mixed phase and on the Isospin  Distillation are
sensitive to the symmetry term in the hadron sector, although the main 
isospin effects are present for all the parametrizations of the isovector 
interaction. At variance,
  for the quark 
sector the lack of explicit isovector
terms could strongly affect the location 
of the phase transition in asymmetric matter and the related 
expected observables.

In conclusion the aim of this work is twofold:
\begin{itemize}
\item
{To stimulate new experiments on isospin effects in heavy ion collisions 
at intermediate energies (in a few $AGeV$ range) with  attention
to the isospin content of produced particles and to elliptic flow properties,
in particular for high-$p_t$ selections.}
\item{To stimulate more refined models of effective Lagrangians for
non-perturbative QCD, where isovector channels are consistently accounted for
 and/or gluon correlations, leading to diquark condensates, can induce 
symmetry 
energy effects.}
\end{itemize}

\subsection*{Acknowledgments}

This project is supported by the National Natural Science Foundation of China
under Grant Nos. 10875160, 11075037 and the INFN of Italy. 
V.B. is grateful for the warm
hospitality at Laboratori Nazionali del Sud, INFN. This collaboration is
supported in part by the Romanian Ministry for Education and Research under 
the CNCSIS contract PNII ID-946/2007.

\vspace{0.5cm}

\appendix

\section{Equation of state for hadronic matter}

\subsection*{A1. Nonlinear (NL) relativistic mean field model with constant 
couplings}

A Lagrangian density with ``minimal'' meson channels and non-linear terms
is used. The nuclear interaction is mediated by two isoscalar,
the scalar $\sigma$ and the vector $\omega$, and two isovector, 
 the scalar $\delta$ and the vector $\rho$, mesons. Non linear terms are 
considered only for the $\sigma$ contribution to account for the correct
compressibility around saturation. Constant nucleon-meson couplings are used,
chosen to reproduce the saturation properties and to represent a reasonable
average of the density dependence predicted by Relativistic 
Dirac-Brueckner-Hartree-Fock  (DBHF) calculations 
\cite{hofmann01,goegelein08}, see details in refs.\cite{liubo02,baranPR}.

\noindent
\begin{eqnarray}\label{lagrdens}
{L} &=& \bar{\psi}[i\gamma_{\mu}\partial^{\mu}-
(M-g_{\sigma}\sigma -g_{\delta}\vec{\tau}\cdot\vec{\delta})
\nonumber\\&&
-g{_\omega}\gamma_\mu\omega^{\mu}-g_\rho\gamma^{\mu}\vec\tau\cdot
\vec{b}_{\mu}]\psi \nonumber \\&&
+\frac{1}{2}(\partial_{\mu}\sigma\partial^{\mu}\sigma-m_{\sigma}^2\sigma^2)
-U(\sigma)\nonumber\\&&
+\frac{1}{2}m^2_{\omega}\omega_{\mu} \omega^{\mu}
+\frac{1}{2}m^2_{\rho}\vec{b}_{\mu}\cdot\vec{b}^{\mu} \nonumber
\\&&
+\frac{1}{2}(\partial_{\mu}\vec{\delta}\cdot\partial^{\mu}\vec{\delta}
-m_{\delta}^2\vec{\delta^2})\nonumber\\&&
 -\frac{1}{4}F_{\mu\nu}F^{\mu\nu}
-\frac{1}{4}\vec{G}_{\mu\nu}\vec{G}^{\mu\nu},
\end{eqnarray}

\noindent
where
$F_{\mu\nu}\equiv\partial_{\mu}\omega_{\nu}-\partial_{\nu}\omega_{\mu}$,
$\vec{G}_{\mu\nu}\equiv\partial_{\mu}\vec{b}_{\nu}-\partial_{\nu}
\vec{b}_{\mu}$,
and the $U(\sigma)$ is the nonlinear potential of $\sigma$ meson :
$U(\sigma)=\frac{1}{3}a\sigma^{3}+\frac{1}{4}b\sigma^{4}$.

The EoS for nuclear matter
 at finite temperature in the mean-field approximation (RMF) is given by
the energy density
\noindent
\begin{eqnarray}\label{edens}
&&\epsilon= 2 \sum_{i=n,p}\int \frac{{\rm
d}^3k}{(2\pi)^3}E_{i}^*(k) (f_{i}(k)+\bar{f}_{i}(k))\nonumber\\&&
+\frac{1}{2}m_\sigma^{2}\sigma^2 + U(\sigma) \nonumber \\&&
+\frac{1}{2}\frac{g_\omega^2}{m_\omega^{2}}\rho_B^2
+\frac{1}{2}\frac{g_\rho^2}{m_\rho^{2}}\rho_3^2
+\frac{1}{2}\frac{g_\delta^2}{m_\delta^{2}}\rho_{s3}^{2}~,
\end{eqnarray}

and pressure

\noindent
\begin{eqnarray}\label{press}
&& p =\frac{2}{3}\sum_{i=n,p}\int \frac{{\rm d}^3k}{(2\pi)^3}
\frac{k^2}{E_{i}^*(k)} (f_{i}(k)+\bar{f}_{i}(k))\nonumber\\&&
-\frac{1}{2}m_\sigma^2\phi^2
-U(\phi) \nonumber \\&&
+\frac{1}{2}\frac{g_\omega^2}{m_\omega^{2}}\rho_B^2
+\frac{1}{2}\frac{g_\rho^2}{m_\rho^{2}}\rho_3^2
+\frac{1}{2}\frac{g_\delta^2}{m_\delta^{2}}\rho_{s3}^{2}~,
\end{eqnarray}

where ${E_i}^*=\sqrt{k^2+{M_i^*}^2}$. The nucleon effective masses
are defined as

\noindent
\begin{equation}\label{mstar}
{M_i}^*=M-g_{\sigma}\sigma \mp g_{\delta}\delta_3~~~ (-~proton, +~neutron).
\end{equation}

The field equations in the relativistic mean field (RMF) approach are

\begin{eqnarray}\label{fields}
&&\sigma=-\frac{a}{m_\sigma^2}\sigma^2-\frac{b}{m_\sigma^2}\sigma^3
+\frac{g_\sigma}{m_\sigma^2}(\rho_{sp}+\rho_{sn})~,\\&&
\omega_0=\frac{g_\omega}{m_\omega^2}\rho~,\\&&
b_0=\frac{g_\rho}{m_\rho^2}\rho_3~,\\&&
\delta_3=\frac{g_\delta}{m_\delta^2}(\rho_{sp}-\rho_{sn})~,
\end{eqnarray}

with the baryon density $\rho\equiv \rho_{B}^{H}=\rho_{p}+\rho_{n}$
and $\rho_3^H=\rho_p-\rho_n$,
$\rho_{sp}$ and $\rho_{sn}$ are the scalar densities for proton and neutron, 
respectively.
The $f_i(k)$ and $\bar{f}_{i}(k)$ in Eqs.(\ref{edens},\ref{press}) 
are the fermion and 
antifermion
distribution functions for protons and neutrons ($i=p,n$):

\noindent
\begin{eqnarray}\label{ffbar}
&&f_i(k)=\frac{1}{1+\exp\{({E_i}^*(k)-{\mu_i^*})/T \} }\,,\\&&
\bar{f}_{i}(k)=\frac{1}{1+\exp\{({E_i}^*(k)+{\mu_i^*})/T \} }.
\end{eqnarray}

where the effective chemical potentials ${\mu_{i}}^*$
is determined by the nucleon densities $\rho_{i}$

\begin{equation}\label{bardens}
\rho_i=2\int\frac{{\rm d}^3k}{(2\pi)^3}(f_{i}(k)-\bar{f}_{i}(k))\,,
\end{equation}

while the scalar densities $\rho_{s,i}$, which gives the coupling to the 
scalar fields
are given by
\begin{eqnarray}\label{rhoscal}
\rho_{s,i}=2 \int\frac{{\rm d}^3k}{(2\pi)^3}\frac{M_{i}^*}{E_{i}^*}
(f_{i}(k)+ \bar{f}_{i}(k))~,
\end{eqnarray}
 note the ${M_i^*}/{E_i^*}$ quenching factor at high baryon density.
Clearly at zero temperature the $\mu_i^*$ reduce to the in medium
Fermi energies ${E_{Fi}}*=\sqrt{k_{Fi}^2+{M_i^*}^2}$.

The $\mu_i^*$ are related to the chemical potentials
$\mu_i=\partial\epsilon/\partial\rho_i$
in terms
of the vector meson mean fields by the equation

\begin{eqnarray}\label{npchempot}
&&\mu_i= \mu_i^*+ \frac{g_{\omega}^2}{m_\omega^2}\rho
\mp \frac{g_{\rho}^2}{m_{\rho}^2}\rho_3
\nonumber\\
&&~~(i=n,p: -~neutron, +~proton)
\end{eqnarray}

The baryon and isospin chemical potentials in the hadron phase
can be expressed in terms of the (p,n) ones
as 
\begin{eqnarray}\label{hchem}
\mu_B^{H} = \frac{\mu_p + \mu_n}{2}, ~~~~~
\mu_3^{H} = \frac{\mu_p - \mu_n}{2}~.
\end{eqnarray}

In presence of the coupling to the two isovector  $\rho,\delta$-meson fields,
 the expression for the symmetry energy has a simple transparent
form, see \cite{liubo02,grecoPRC03,baranPR}:

\begin{equation}\label{esym}
E_{sym}(\rho)=\frac{1}{6} \frac{k_{F}^{2}}{E_{F}^*}+
\frac{1}{2}[f_{\rho}-f_{\delta}(\frac{M^*}{E_{F}^*})^{2}]\rho~,
\end{equation}

where $M^*=M-g_{\sigma}\sigma$ and
${E_F}^*=\sqrt{k_{F}^2+{M^*}^2}$.

Now we easily see that in the $NL\rho\delta$ choice we have a large increase 
of the symmetry energy at high baryon densities. The potential simmetry term
is given by the combination 
$[f_{\rho}-f_{\delta}({M^*}/{E_{F}^*})^{2}]$ of the repulsive 
vector $\rho$ and attractive scalar $\delta$ isovector couplings. Thus, 
when the $\delta$ is included we have to increase the $\rho$-meson coupling 
in order to reproduce the same asymmetry parameter $a_4$ at saturation. 
The net effect will be a stiffer symmetry energy at higher baryon densities 
due to the
$M^*/E_F^*$ quenching of the attractive part. Of course
this mechanism can be largely modified if some density dependece is
explicitly included in the meson-nucleon couplings, as we will see in 
the $DDRH$
forces (Appendix A2), also used in this paper..

\subsection*{Parameter determination}

The coupling constants
are fixed from good 
saturation properties and from averaged Dirac-Brueckner-Hartree-Fock 
estimations, see 
the detailed
discussions in refs.\cite{liubo02,grecoPRC03,baranPR}. 
DBHF indications of a density dependence of the meson-nucleon couplings
at high baryon densities will be accounted for in the $DDRH$ forces
of the next Subsection A2.

The isoscalar part of 
the EoS is chosen to be rather soft at high densities, see \cite{liubo05},
in order to satisfy the experimental constraints from collective flows and
kaon production in intermediate energy heavy ion collisions
\cite{daniel02,fuchs06}.

The coupling constants, $f_{i}\equiv g_{i}^{2}/m_{i}^{2}$,
$i=\sigma, \omega, \rho, \delta$, and the two parameters of the $\sigma$
self-interacting terms : $A\equiv a/g_{\sigma}^{3}$ and $B\equiv
b/g_{\sigma}^{4}$ are reported in Table 1. The $\sigma $ mass is fixed at
$550MeV$.
The corresponding properties of nuclear matter are listed in Table 2.
Here the binding energy is defined $E/A=\epsilon/\rho-M$.

\par
\vspace{0.3cm}
\noindent

\begin{center}
{{\large \bf Table 1.}~Parameter set.}
\par
\vspace{0.5cm}
\noindent

\begin{tabular}{c|c|c} \hline
$Parameter~~Set$  &NL$\rho$  &NL$\rho\delta$ \\ \hline
$f_\sigma~(fm^2)$  &10.32924  &10.32924    \\ \hline
$f_\omega~(fm^2)$  &5.42341   &5.42341     \\ \hline
$f_\rho~(fm^2)$    &0.94999   &3.1500      \\ \hline
$f_\delta~(fm^2)$  &0.000     &2.500       \\ \hline
$A~(fm^{-1})$      &0.03302   &0.03302     \\ \hline
$B$                &-0.00483  &-0.00483    \\ \hline
\end{tabular}
\end{center}
\vspace{0.5cm}
\noindent

\begin{center}

{{\large \bf Table 2.}~Saturation properties of nuclear matter.}

\par
\vspace{0.5cm}

\noindent
\begin{tabular}{ c c c } \hline
$\rho_{0}~(fm^{-3})$ &0.16   \\ \hline
$E/A ~(MeV)$         &-16.0  \\ \hline
$K~(MeV)$            &240.0   \\ \hline
$E_{sym}~(MeV)$      &31.3    \\ \hline
$M^{*}/M $           &0.75    \\ \hline
\end{tabular}
\end{center}

We finally note that these Lagrangians have been already used for flow
\cite{greco03}, pion production \cite{gaitanos04}, isospin tracer
\cite{gaitanosplb04} and kaon production \cite{ferini06} calculations 
for relativistic heavy ion collisions with
an overall good agreement to data.

\subsection*{A2. DDRH forces: Relativistic Mean Field model with
density dependent couplings}

The ``minimal'' Lagrangian density has the same form of
the Eq.(\ref{lagrdens}), now with density dependent couplings and of course
without non-linear terms (the $U(\sigma)$ potential). Apart the effect of an 
explicit variation of the meson-nucleon couplings with baryon density we will 
expect new terms in the variational derivative of the Lagrangian density,
the rearrangement terms $\Sigma^R_\mu$ that will affect the nucleon field 
equation as well as the energy-momentum tensor and so the $EoS$ and the
 nucleon chemical potentials,
see \cite{fuchs95,typel99,liubo07}.
The nucleon field equation in a mean field approximation ($RMF$) is
\begin{eqnarray}\label{eq:6}
&& (i\gamma_{\mu}\partial^{\mu}-(M- g_{\sigma}\sigma
-g_\delta{\tau_3}\delta_3)-g_\omega\gamma^{0}{\omega_0}\nonumber\\
&& -g_\rho\gamma^{0}{\tau_3}{b_0}
+\gamma^0\Sigma_0^R)\psi=0
\end{eqnarray}

\noindent
with

\begin{eqnarray}\label{DDfields}
&& \sigma=\frac{g_\sigma}{m_{\sigma}^2} \rho_{s}
=\frac{g_\sigma}{m_{\sigma}^2} (\rho_{sp}+\rho_{sn}), \nonumber \\
&& \omega_{0}=\frac{g_\omega}{m^2_{\omega}} <\bar\psi{\gamma^0}\psi>
=\frac{g_\omega}{m_{\omega}^2}\rho
=\frac{g_\omega} {m_{\omega}^2}(\rho_p+\rho_n),\nonumber \\
&& b_{0}=\frac{g_\rho}{m^2_{\rho}} <\bar\psi{\gamma^0}\tau_3\psi>
=\frac{g_\rho}{m^2_{\rho}}\rho_3,\nonumber \\
&& \delta_3=\frac{g_\delta}{m^2_{\delta}}<\bar\psi\tau_3\psi>
=\frac{g_{\delta}}{m^2_{\delta}} \rho_{s3}
\end{eqnarray}

and the rearrangement term

\begin{eqnarray}\label{rearr}
\Sigma_0^{R}=(\frac{\partial g_\sigma}{\partial \rho})
\frac{g_\sigma}{m_\sigma^2} \rho_s
+(\frac{\partial g_\delta}{\partial \rho}) \frac{g_\delta}{m_\delta^2}
\rho_{s3}^2 \nonumber\\
-(\frac{\partial g_\omega}{\partial \rho}) \frac{g_{\omega}}{m_{\omega}^2} 
\rho^2
-(\frac{\partial g_\rho}{\partial \rho}) \frac{g_{\rho}}{m_{\rho}^2} 
\rho_{3}^{2},
\end{eqnarray}

\noindent
where
$\rho_3=\rho_p-\rho_n$ and $\rho_{s3}=\rho_{sp}-\rho_{sn}$,  with 
$\rho_i$, $\rho_{si}$ (i=n,p)
 the nucleon and the scalar densities, see Subsection A1.


Neglecting the derivatives of meson fields, the energy-momentum
tensor in $RMF$ approximation is given by

\begin{eqnarray}\label{eq:8}
T_{\mu\nu}=i\bar{\psi}\gamma_{\mu}\partial_{\nu}\psi+[\frac{1}{2}
 m_{\sigma}^2\sigma^2
 -\frac{1}{2}m^2_{\omega}\omega_{\lambda} \omega^{\lambda} \nonumber \\
-\frac{1}{2}m^2_{\rho}\vec{b_{\lambda}}\vec{b^{\lambda}}
+\frac{1}{2}m_{\delta}^2 \vec{\delta} ^2+\bar{\psi}\Sigma_{\lambda}^{R}
\gamma^\lambda\psi]g_{\mu\nu}.
\end{eqnarray}


The equation of state ($EOS$) for nuclear matter at finite temperature 
can be obtained from
the thermodynamic potential.
Using the above meson field equations \ref{DDfields}, the 
energy density has the form

\begin{eqnarray}\label{eq:13}
&&\epsilon=\sum_{i=n,p}{2}\int \frac{{\rm d}^3k}{(2\pi)^3}E_{i}^*(k)
(f_i(k)+\bar{f}_{i}(k))
+\frac{1}{2}\frac{g_{\sigma}^2}{m_\sigma^2}\rho_s^2 \nonumber\\
&& +\frac{1}{2} \frac{g_{\omega}^2}{m_\omega^2}\rho^2
+\frac{1}{2}\frac{g_{\rho}^2}{m_{\rho}^2}\rho_3^2
+ \frac{1}{2}\frac{g_{\delta}^2}{m_{\delta}^2}\rho_{s3}^2,
\end{eqnarray}

and the pressure

\begin{eqnarray}\label{eq:14}
&& p =\sum_{i=n,p} \frac{2}{3}\int \frac{{\rm d}^3k}{(2\pi)^3}
\frac{k^2}{E_{i}^*(k)}
(f_i(k)+\bar{f}_{i}(k))
-\frac{1}{2}\frac{g_{\sigma}^2}{m_\sigma^2}\rho_s^2 \nonumber\\
&& +\frac{1}{2}\frac{g_{\omega}^2}{m_\omega^2}\rho^2
+\frac{1}{2}\frac{g_{\rho}^2}{m_{\rho}^2}\rho_3^2
-\frac{1}{2}\frac{g_{\delta}^2}{m_{\delta}^2}\rho_{s3}^2-\Sigma_{o}^{R}\rho.
\end{eqnarray}

The nucleon chemical potentials $\mu_i$ are given in terms of the
vector meson mean fields as in the constant coupling case, 
Eq.(\ref{npchempot}), apart the new rearrangement term

\begin{eqnarray}\label{eq:15}
&&\mu_i= \mu_i^*+ \frac{g_{\omega}^2}{m_\omega^2}\rho
\mp \frac{g_{\rho}^2}{m_{\rho}^2}\rho_3
-\Sigma_{o}^{R}\ \nonumber\\
&&~~(i=n,p: -~neutron, +~proton)
\end{eqnarray}

with the effective masses related to the scalar fields
as before

\begin{eqnarray}\label{eq:11}
&&{M_i}^*=M-\frac{g_{\sigma}^2}{m_\sigma^2}\rho_s
\mp \frac{g_{\delta}^2}{m_{\delta}^2}\rho_{s3}\nonumber\\
&&~~(i=n,p: +~neutron, -~proton)
\end{eqnarray}

\begin{figure}
\begin{center}
\includegraphics[scale=0.33]{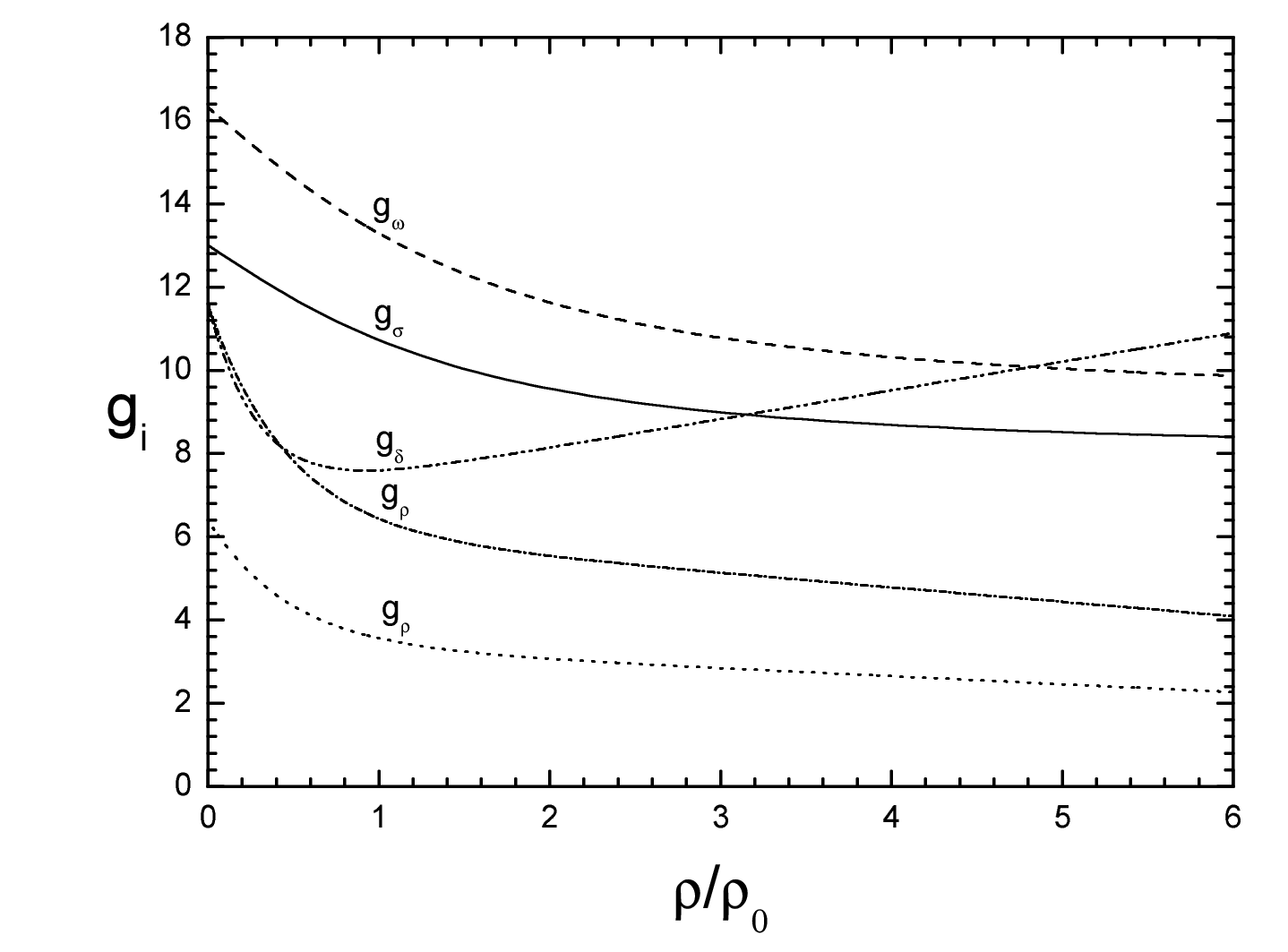}
\caption{
Density dependence of the meson-nucleon couplings in the used $DDRH$ 
interactions. The $g_\rho$ dotted line at the bottom corresponds to 
the $DDRH\rho$
case, without the $\delta$ meson.
}
\label{couplings}
\end{center}
\end{figure}

\subsection*{Density dependence parametrization}

A general form of parametrization for the density dependence of the
meson-nucleon couplings can be given by :

\begin{equation}\label{eq:31}
g_{i}(\rho)=g_{i}(\rho_0)f_{i}(x), ~~~~for ~~i=\sigma,\omega,\rho,\delta,
\end{equation}

\noindent
with $x=\rho/\rho_0$ and $\rho_0$ saturation density. As already mentioned
the $f_{i}(x)$ are chosen in order to reproduce the density dependence of 
the couplings deduced from microscopic DBHF calculations. For symmetric matter
this analysis has been performed in ref.\cite{typel99} using for the 
isoscalar mesons a functional
of the form 

\begin{equation}\label{eq:32}
f_{i}(x)=a_{i}\frac{1+b_{i}(x+d_i)^2}{1+c_{i}(x+d_i)^2},~~~~ 
for~~i=\sigma,\omega,
\end{equation}

\noindent

In the case of asymmetric matter the following parametrization has been 
proposed for the isovector couplings \cite{avancini04} :

\begin{equation}\label{eq:33}
f_{i}(x)=a_{i}exp[-b_{i}(x-1)] - c_{i}(x-d_i), ~~~~ for~~ i=\rho,\delta.
\end{equation}

\noindent
in this way it easier to reproduce the important difference of the 
$\delta,~\rho$ couplings at high density.

We follow the two suggestions, the parametrization form and parameters 
of our $DDRH$ forces are taken from ref.\cite{typel99}
for $\sigma$, $\omega$ mesons and ref.\cite{avancini04} for $\rho$, $\delta$ 
mesons, respectively.
 All parameters  are listed in Table 3.
 The density dependent couplings as a function of baryon density are displayed 
in Fig.\ref{couplings}.


\vspace{0.3cm}
\noindent

\begin{center}
{{\large \bf Table 3.}~~DDRH~Parameters}.
\par
\vspace{0.5cm} \noindent

\begin{tabular}{c|c|c|c|c|c} \hline
 Model  &\multicolumn{2}{c|} {$DDRH$} &DDRH$\rho$  &\multicolumn{2}{c}{DDRH$\rho\delta$} \\ \hline
 Meson          &$\sigma$   &$\omega$   &$\rho$           &$\rho$    &$\delta$ \\ \hline
$m_i~(MeV)$     &550        &783        &770              &770       &980      \\ \hline
$g_i(\rho_{0})$ &10.73   &13.29   &3.59            &6.48     &7.59     \\ \hline
$a_i$           &1.37   &1.40   &0.095         &0.095  &0.02   \\ \hline
$b_i$           &0.23   &0.17   &2.17            &2.17     &3.47    \\ \hline
$c_i$           &0.41   &0.34   &0.05          &0.05   &-0.09   \\ \hline
$d_i$           &0.90   &0.98   &17.84          &17.84   &-9.81    \\ \hline
\end{tabular}
\end{center}

\vspace{0.3cm}

The choice of the $g_i(\rho_0)$ couplings at saturation is performed in order
to have the same nuclear matter normal properties of the Non Linear RMF models
(Table 2) of the Subsection A1. The EoS of symmetric matter at high density 
is also not affected, as we can see comparing the binodal surfaces for zero 
asymmetry of Figs.\ref{tcrit}, \ref{tcritDD} (solid lines).  

At variance,
 the different behavior of the isovector couplings at high density, increase 
of $g_\delta$ and descrease of $g_\rho$, will
contribute to get a much softer symmetry energy at high baryon densities,
 Fig.\ref{esymDD}.
It is easy to check that in this way in nucleonic models of neutron stars
the proton fraction limit for the onset of direct URCA processes is hardly 
reached, see the analysis of ref.\cite{klaehn06}. For the purpose of the 
present paper the use of $DDRH$ interactions is important in order to show that
the expected isospin effects on the mixed phase are present even with
much softer symmetry terms at high baryon densities.

\section{Nambu-Iona Lasinio model for asymmetric matter} 

From the above discussion it appears (extremely) important to include the 
Isospin degree of freedom in any effective QCD dynamics. 
A first approach can be supplied by a two-flavor Nambu-Jona Lasinio 
model where the isospin asymmetry can be included in a flavor-mixing picture 
\cite{frank03,buballa05}. 
The lagrangian is given by
\begin{equation}
{ L} = {L}_0 + {L}_1 + {L}_2 , 
\end{equation}
with ${ L}_0$ the free part 
$$ 
{ L}_0 = {\bar \psi} ( i\not \!\partial - m ) \psi, 
$$
and the two different interaction part given by
\begin{eqnarray}
{L}_1 = G_1 \Big\{ ({\bar \psi}
\psi)^2 + ({\bar \psi}\,\vec\tau \psi)^2 + ({\bar \psi}\,i\gamma_5 \psi)^2 + 
({\bar
\psi}\,i\gamma_5\vec\tau \psi)^2 \Big\} \nonumber \\
{L}_2 = 
 G_2 \Big\{ ({\bar \psi} \psi)^2 - ({\bar \psi}\,\vec\tau \psi)^2 - ({\bar
\psi}\,i\gamma_5 \psi)^2 + ({\bar \psi}\,i\gamma_5\vec\tau \psi)^2 \Big\}.
\end{eqnarray}
In the mean field approximation the new Gap Equations are 
$M_i=m_i-4G_1 \Phi_i-4G_2 \Phi_j$, $i \not= j,(u,d)$ , where the 
$\Phi_{u,d}=<\bar u u>,<\bar d d>$ are the two (negative) condensates 
which are given by
\begin{equation}
\Phi_f = - 2N_c \int \frac{d^3p}{(2 \pi)^3} \,
\frac{M_f}{E_{p,f}}\Big\{ 1 - f^-(T,\mu_f) 
- f^+(T,\mu_f) \Big\}.
\end{equation}
and $m_{u,d}=m$
the (equal) current masses. 

Introducing explicitily a flavor mixing, i.e.
the dependence of the constituent mass of a given flavor to both condensate,
via $G_1=(1-\beta) G_0, G_2= \beta G_0$ we have the coupled equations
\begin{eqnarray}
&&M_u=m - 4 G_0 \Phi_u + 4 \beta G_0 (\Phi_u - \Phi_d), \nonumber \\ 
&&M_d=m - 4G_0 \Phi_u + 4 (1-\beta) G_0 (\Phi_u - \Phi_d).
\label{mixing}
\end{eqnarray}
For $\beta=1/2$ we have back the usual NJL ($M_u=M_d$), while small/large
mixing is for $\beta \Rightarrow 0$/$\beta \Rightarrow 1$ respectively.
The value of $\beta$ has a consequence on the structure 
of the phase diagram in the region of low temperatures 
and high chemical potential. In fact as shown in \cite{frank03,buballa05} 
for $\beta = 0$ there are two distinct phase transitions 
for the up quarks and for the down quarks, but for this 
value the interaction is symmetric under $U_A(1)$ 
transformations and it is unrealistic. While for 
$\beta \geq 0.1$ the $U_A(1)$ symmetry becomes explicitly
broken and there is only a single first order phase transition.
Realistic estimations of $\beta$ fitting the physical $\eta$-meson  mass
give a value of $\beta \approx 0.11$  \cite{frank03,shao06}.


In neutron rich matter $\mid \Phi_d \mid$ decreases more rapidly due to the 
larger $\rho_d$ and so $(\Phi_u -\Phi_d)<0$. In the ``realistic'' small mixing 
case we will get a definite $M_u>M_d$ 
splitting at high baryon density (before the chiral restoration). 
This expectation is confirmed by a full calculation of the coupled gap 
equations 
with standard parameters  \cite{erice08,plum_tesi}. 
All that can indicate a more fundamental confirmation of the $m^*_p>m^*_n$ 
splitting in the hadron phase,
as suggested by the effective $QHD$ model with the isovector scalar $\delta$ 
coupling, see \cite{liubo02,baranPR}.

However such isospin mixing effect results in a very small variation
of the symmetry energy in the quark phase, still related only to the Fermi
kinetic contribution. 
 In fact this represents just a very first step towards a more complete
treatment of isovector contributions in effective quark models, of large 
interest for the discussion of the phase transition at high densities. 




\begin{thebibliography}{99}

\bibitem{muller}H.  M$\ddot{u}$ller,  Nucl.Phys. {\bf A618} 
349 (1997).

\bibitem{ditoro} M. Di Toro, A. Drago, T. Gaitanos, V. Greco, A. Lavagno,
 Nucl.Phys. {\bf A775} 102 (2006).

\bibitem{erice08} 
M. Di Toro et al. Progr.Part.Nucl.Phys. {\bf 62} 389-401 (2009).

\bibitem{fair}
P. Senger  et al., J.Phys.G:Nucl.Part.Phys. {\bf 36} 064037 (2009).

\bibitem{nica}
See the Website $<http://nica.jinr.ru/>$.

\bibitem{burgio02}
G.F. Burgio, M. Baldo, P.K. Sahu, H.-J. Schulze, 
 Phys.Rev. {\bf C66} 025802 (2002).

\bibitem{haensel05}
M. Bejger, P. Haensel, J.L. Zdumik,
 Monthly Notices of the Royal Astronomy Society {\bf 359} 699 (2005). 

\bibitem{nicotra06}
O.E. Nicotra, M. Baldo, G.F. Burgio, H.-J. Schulze,
Phys.Rev. {\bf D74} 123001 (2006).

\bibitem{baldo07}
M.Baldo, G.F. Burgio, P.Castorina, S. Plumari, D. Zappala',
Phya.Rev. {\bf C66} 035804 (2007).

\bibitem{burgio08}
G.F. Burgio, S. Plumari, 
 Phys.Rev. {\bf D77} 085022 (2008).

\bibitem{providencia09}
A. Rahbi, H.Pais, P.K. Panda, C. Procidencia,
{\em Quark-hadron phase transition in a neutron star with strong magnetic 
 fields}, arXiv:0909.1114[nucl-th].

\bibitem{baranPR} V. Baran, M. Colonna, V. Greco, M. Di Toro,
  Phys.Rep. {\bf 410} 335 (2005).
 
\bibitem{fuwo06}
C. Fuchs, H.H. Wolter,
 Eur. Phys. J. {\bf A30} 5 (2006).

\bibitem{baoPR}
B.A. Li, L.W. Chen, C.M. Ko, Phys.Rep. {\bf 465} 113 (2008).

\bibitem{trautmann10}
W. Trautmann,
Nucl. Phys. {\bf A834} 548c (2010) and arXiv:1001.3867[nucl-exp].

\bibitem{giordano10}
V. Giordano, M. Colonna, M. Di Toro, V. Greco, J. Rizzo,
 Phys.Rev. {\bf C81} 044611 (2010).

\bibitem{ditoro10}
M. Di Toro, V. Baran, M. Colonna, V. Greco,
J. Phys.G: Nucl.Part.Phys. {\bf 37} 083101 (2010).


\bibitem{MIT}  A. Chodos et al., Phys.Rev. {\bf D9} 3471 (1974).


\bibitem{NJL}
Y. Nambu and G. Jona-Lasinio, 
Phys.Rev. {\bf 122} 345 (1961); {\bf 124} 246 (1961).

\bibitem{buballa05}
M. Buballa, Phys. Rep. {\bf 407} 205 (2005)


\bibitem{pagliara10}
G. Pagliara, J. Schaffner-Bielich,
{\it Phase transition from nuclear matter to color superconducting
 quark matter: the effects of isospin}, arXiv:1003.1017[nucl-th].




\bibitem{SW85} B.D. Serot and J.D. Walecka, {\it Adv.\  Nucl.\ Phys.} 
{\bf 16} 1 (1985).

\bibitem{liubo02} B. Liu, V. Greco, V. Baran, M. Colonna, M. Di Toro,
  Phys.Rev. {\bf C65} 045201 (2002).

\bibitem{page06}
D. Page, S. Reddy, 
Ann.Rev.Nucl.Part.Sci. {\bf 56} 327 (2006). 

\bibitem{daniel02}
P. Danielewicz, R. Lacey, W.G. Lynch, 
Science {\bf 298} 1592 (2002).

\bibitem{fuchs06}
C. Fuchs,
Prog.Part.Nucl.Phys. {\bf 56} 1-103 (2006).

\bibitem{bmuller95}
B. M$\ddot{u}$ller,
Rep.Prog.Phys. {\bf 58} 611 (1995).

\bibitem{Landaustat} 
 L.D. Landau and L. Lifshitz,
{\em Statistical Physics}
 Pergamon Press 1969, Oxford.

\bibitem{greco03}
V. Greco et al., Phys.Lett. {\bf B562} 215 (2003).

\bibitem{gaitanos04}
T. Gaitanos et al., Nucl.Phys. {\bf A732}  24 (2004).

\bibitem{ferini06}
G. Ferini, T. Gaitanos, M. Colonna, M. Di Toro. H.H. Wolter,
Phys.Rev.Lett. {\bf 97} 202301 (2006).

\bibitem{GlendenningPRL18} 
N.K. Glendenning, S.A. Moszkowski,  
Phys.Rev.Lett. {\bf 67} 2414 (1991).

\bibitem{vretenar03}
D. Vretenar, T. Niksic, P. Ring,
 Phys.Rev. {\bf C68} 024310 (2003), and refs. therein.




\bibitem{baran98}
V. Baran, M. Colonna, M. Di Toro, A. Larionov,
Nucl.Phys. {\bf A632} 287-303 (1998).

\bibitem{chomazPR}
P. Chomaz, M. Colonna, J. Randrup.
Phys.Rep. {\bf 389} 263-440 (2004).

\bibitem{Qasy}
 In principle there is nothing wrong with $\alpha^Q>1$ evaluations
(although we note that for pure "neutron" matter the quark phase 
asymmetry should be 1, since $\rho_d=2\rho_u$). Indeed for low $\chi$-values,
$\chi<0.1$, very small quark concentrations, we can get $\alpha^Q$ values 
slightly 
larger than 1. However we are cautious about these results since we can 
expect also some numerical problems. In fact for very small $\chi$ values
the weight of the $\alpha^Q$ contribution in the minimization procedure 
is expected not too relevant, as we can clearly see from the 
Eq.(\ref{alphamix}). 
In any case the important point is that this is not affecting the 
general discussion about the isospin distillation.


\bibitem{sissakian08}
A.N. Sissakian, A.S. Sorin, V.D. Toneev,
Phys.Part.Nucl. {\bf 39} 1062 (2008).

\bibitem{kogut04}
J.B. Kogut, D.K. Sinclair,
 Phys.Rev. {\bf D70} 095401 (2004).

\bibitem{toublan05}
D. Toublan, J.B. Kogut, 
Phys.Lett. {\bf B605} 129 (2005).




\bibitem{fuchs95}
C. Fuchs, H. Lenske, H.H. Wolter,
Phys.Rev. {\bf C52} 3043 (1995). 

\bibitem{typel99}
S. Typel, H.H. Wolter,
Nucl.Phys. {\bf A656} 331 (1999).

\bibitem{avancini04} 
S.S. Avancini, L. Brito, D.P. Menezes, C. Providencia,
Phys.Rev. {\bf C70} 015203 (2004).

\bibitem{hofmann01}
F. Hofmann, C.M. Keil, H. Lenske, 
Phys.Rev. {\bf C64} 034314 (2001).

\bibitem{goegelein08}
P. Goegelein, E.N.E. van Dalen, C.  Fuchs, H. M$\ddot{u}$ther,  
Phys.Rev. {\bf C77} 025802 (2008).

\bibitem{liubo07}
B. Liu et al., Phys.Rev. {\bf C75} 048801 (2007).





\bibitem{frank03}
M. Frank, M. Buballa, M. Oertel, 
 Phys.Lett. {\bf B562} 221 (2003).



\bibitem{shao06}
Guo-yun Shao et al., 
 Phys.Rev. {\bf D73} 076003 (2006).

\bibitem{plum_tesi}
S. Plumari, Ph.D.Thesis, 2009  Univ.Catania.

\bibitem{kaiser09}
N. Kaiser, W. Weise, 
Phys.Lett. {\bf B671} 25 (2009).

\bibitem{weise07}
W. Weise,
Prog.Theor.Phys.Suppl. {\bf 170} 161 (2007).

\bibitem{alford08}
M.G. Alford, A. Schmitt, K. Rajagopal, T. Schafer,
Rev.Mod.Phys. {\bf 80} 1455 (2008).


\bibitem{huang03}
Mei Huang, Pengfei Zhuang, Weiqin Chao,
Phys.Rev. {\bf D67} 065015 (2003).

\bibitem{horst}
H. St$\ddot{o}$ecker, private communication.

\bibitem{ferini05}
G. Ferini, M. Colonna, T. Gaitanos, M. Di Toro,
 Nucl.Phys. {\bf A762} 147 (2005).

\bibitem{csernai99}
L. Csernai, D. Rohrich, 
Phys.Lett. {\bf B458}  454 (1999).

\bibitem{fries08}
R.J. Fries, V. Greco V, P. S$\ddot{o}$rensen, 
Ann.Rev.Nucl.Part.Sci. {\bf 58} 177 (2008).

\bibitem{grecoPRC03} 
V. Greco, M. Colonna, M. Di Toro, F. Matera,
 Phys.Rev. {\bf C67} 015203 (2003).

\bibitem{liubo05}
B. Liu, H. Guo, M. Di Toro, V. Greco, 
Eur.Phys.J. {\bf A25} 293 (2005).

\bibitem{gaitanosplb04}
T. Gaitanos, M. Colonna, M. Di Toro, H.H. Wolter,
Phys.Lett. {\bf B595}  209 (2004).
.
\bibitem{klaehn06} 
T. Kl$\ddot{a}$hn et al., Phys.Rev. {\bf C74} 035802 (2006).

\end{thebibliography}
\end{document}